\pdfoutput=1
\PassOptionsToPackage{svgnames}{xcolor}

\documentclass[11pt]{article}

\usepackage{acl}

\usepackage{times}
\usepackage{latexsym}

\usepackage[T1]{fontenc}

\usepackage[utf8]{inputenc}

\usepackage{microtype}

\usepackage{inconsolata}
\usepackage{float}
\usepackage{graphicx}
\usepackage{subcaption}
\usepackage[export]{adjustbox}
\usepackage{wrapfig}  
\usepackage{amsmath}
\usepackage{listings}
\usepackage{xcolor} 
\usepackage{enumitem}
\usepackage{booktabs} 
\usepackage{tcolorbox}
\usepackage{multicol}
\usepackage{courier}
\usepackage[capitalise,nameinlink]{cleveref}
\usepackage{url}
\usepackage{soul}

\definecolor{brown}{rgb}{0.6, 0.4, 0.2}

\usepackage{subcaption}

\definecolor{vgreen}{HTML}{60A917}
\definecolor{vred}{HTML}{CE3A29}
\definecolor{vindigo}{HTML}{4cbbe8}
\definecolor{vblue}{HTML}{00008B}

\newcommand*{\affaddr}[1]{#1}
\newcommand*{\affmark}[1][*]{\textsuperscript{#1}}
\newcommand*{\email}[1]{\texttt{#1}}
\author{
Vipula Rawte\affmark[1]\thanks{\,\,\,Corresponding author.}, Deja Scott\affmark[2], Gaurav Kumar\affmark[2], \\ \bf Aishneet Juneja\affmark[2], \bf Bharat Yaddanapalli\affmark[2], Biplav Srivastava\affmark[1]  \\
\affaddr{\affmark[1]AI Institute, University of South Carolina, USA}\\
\affaddr{\affmark[2]University of South Carolina, USA}\\
\email{\{vrawte@mailbox.,biplav.s@\}sc.edu}
}

\title{Do Voters Get the Information They Want? Understanding Authentic Voter FAQs in the US and How to Improve  for Informed Electoral Participation}

\begin{document}
\maketitle

\begin{abstract}
Accurate information is crucial for democracy as it empowers voters to make informed decisions about their representatives and keeping them accountable. In the US, state election commissions (SECs), often required by law, are the primary providers of Frequently Asked Questions (FAQs) to voters, and secondary sources like non-profits such as  League of Women Voters (LWV) try to complement their information shortfall. However, surprisingly, to the best of our knowledge, there is neither a single source with comprehensive FAQs nor a study analyzing the data at national level to identify current practices and ways to improve the status quo. This paper addresses it by providing the {\bf first  dataset on Voter FAQs  covering all the US states}. Second, we introduce metrics for FAQ information quality score (FIQS)  with respect to  questions, answers, and answers to corresponding questions. Third, we use FIQS to analyze US FAQs to identify  leading, mainstream and lagging content practices and corresponding states. Finally, we identify what states across the spectrum can do to improve FAQ quality and thus,  the overall information ecosystem. Across all 50 U.S. states, 12\% were identified as leaders and 8\% as laggards for FIQS\textsubscript{voter}, while 14\% were leaders and 12\% laggards for FIQS\textsubscript{developer}. The code and sample data are provided at \url{https://anonymous.4open.science/r/election-qa-analysis-BE4E}.

\end{abstract}

\section{Introduction} \label{sec:intro}

\begin{figure}[!ht]
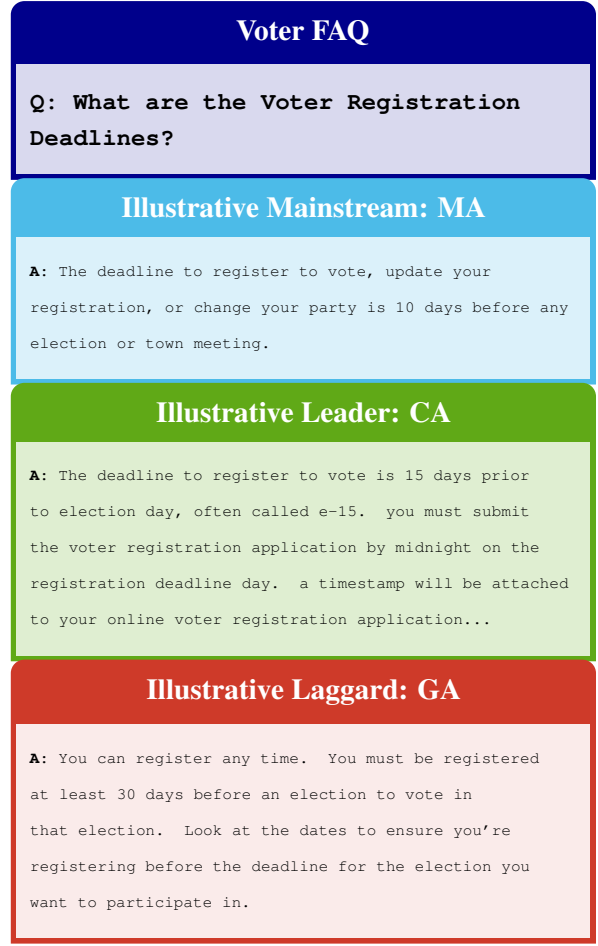

\centering
    
\begin{tcolorbox}[colframe=vblue, colback=vblue!15, coltitle=white, left=0mm, right=0mm, sharp corners=south, boxrule=0.8mm, boxsep=5pt, title=\centering\textbf{Voter FAQ}]
{\bf \ttfamily \footnotesize Q: What are the Voter Registration Deadlines?}
\end{tcolorbox}

\vspace{-1.3em}  

\begin{tcolorbox}[colframe=vindigo, colback=vindigo!20, coltitle=white, left=0mm, right=0mm, sharp corners=south, boxrule=0.8mm, boxsep=5pt, title=\centering\textbf{Illustrative Mainstream: MA}]
{\ttfamily \tiny \tiny \textbf{A:} The deadline to register to vote, update your registration, or change your party is 10 days before any election or town meeting.} 
\end{tcolorbox}

\vspace{-1.3em}  

\begin{tcolorbox}[colframe=vgreen, colback=vgreen!20, coltitle=white, left=0mm, right=0mm, sharp corners=south, boxrule=0.8mm, boxsep=5pt, title=\centering\textbf{Illustrative Leader: CA}]
{\ttfamily \tiny \textbf{A:} The deadline to register to vote is 15 days prior to election day, often called e-15. you must submit the voter registration application by midnight on the registration deadline day. a timestamp will be attached to your online voter registration application...} 
\end{tcolorbox}

\vspace{-1.3em}  

\begin{tcolorbox}[colframe=vred, colback=vred!10, coltitle=white, left=0mm, right=0mm, sharp corners=south, boxrule=0.8mm, boxsep=5pt, title=\centering\textbf{Illustrative Laggard: GA}]
{\ttfamily \tiny \textbf{A:} You can register any time. You must be registered at least 30 days before an election to vote in that election. Look at the dates to ensure you're registering before the deadline for the election you want to participate in.} 
\end{tcolorbox}

\caption{A real-world example of Voter FAQ. Scores of content quality are (FIQS\textsubscript{voter}, FIQS\textsubscript{developer}) - MA (0.41, 0.38);  CA (0.7, 0.7); GA (0.13, 0.18). }
\label{fig:intro-q}
\vspace{-0.2in}
\end{figure}

Democracy is the leading form of governance where people have a say in who governs them. Its success depends on the ability of participants to vote in regular elections and the ability of the government to implement the subsequent orderly transfer of power \cite{american-election-issues,voting-american}.  
Democracy at
a practical level means empowering the voter with a right to choose and providing all relevant and reliable
information including knowing about candidates, campaign finance, voting procedure, processing of
votes, and declaration of results. 
However, around the world, stakeholders are struggling to find accurate information, which is now especially acute in the age of generative Artificial Intelligence (AI) and other technologies from the Natural Language Processing (NLP) and wider AI communities. 

The situation is so bad with  information gap and  disorders that whenever AI is referenced in connection with elections, it often draws negative reactions due to the fear of bots, misinformation, and hacking. 
As a baseline and illustration of the current situation, for elections, OpenAI declared that ChatGPT will defer election questions to human-curated Frequently Asked Questions (FAQs) \cite{openai-chatgpt-elections}, even though it has one of the best performance in QA settings. 
This is particularly disappointing for AI, and especially chatbots, or bots, for short, since they are multi-modal collaborative assistants which have been studied since the early days of AI to help people complete useful tasks. For elections, people could have overcome voting complexity by  accessing authentic information 
conveniently in their own language or words through their smartphones, computers, and home devices.

In the United States (US), state election commissions (SECs), often required by law, are the primary providers of Frequently Asked Questions (FAQs) (see \cref{fig:intro-q}) to voters, and secondary sources like non-profits such as  League of Women Voters (LWV) try to complement their information shortfall. 
However, there is a general perception that  it is hard to find the right, accurate, information and in its absence,  the democratic processes are under increasing threats like {\em information disorders}, a term which covers misinformation, disinformation and malinformation \cite{american-election-issues,ai-maniputation-characterize,bot-politics-misinfo,fake-disinfo-shu2020mining,info-disorder}. 
However, surprisingly, to the best of our knowledge, there is neither a single source with comprehensive FAQs nor a study analyzing the data at national level to identify current practices and ways to improve the status quo. In response, we  provide  a  dataset on Voter FAQs for the NLP community covering all the US states. We next present the related work, followed by data and NLP methods, and then analyze the FAQ data. We use the analysis to identify guidelines that lagging and mainstream states can adopt, and conclude.

In summary, our key contributions are:

\begin{itemize}[leftmargin=15pt] 
    \item We present the \textbf{first NLP dataset of voter FAQs} encompassing all U.S. states (see \cref{sec:data}).
    \item We introduce metrics for FAQ information quality score (FIQS)  with respect to  questions, answers, and answers to corresponding questions (see \cref{sec:setup}).
    \item We use FIQS to analyze US FAQs to identify  leading, mainstream and lagging content practices and corresponding states. (cf. \cref{sec:analysis}).
    \item We identify what states across the spectrum can do to improve FAQ quality and thus,  the overall information ecosystem.  (cf. \cref{sec:guidelines}).
\end{itemize}







    





\section{Related Work} \label{sec:related}

Going beyond studying the negative impacts of  AI on elections, researchers have begun shifting attention to its positive impacts.
AI-driven tools can enhance voter education by offering personalized, real-time responses to common election questions, and they can support policymakers by identifying trends and disparities in voter access or participation. By providing actionable insights, AI could play a transformative role in improving the transparency and efficiency of electoral systems.
In this regard, \cite{srivastava2025vision} outlines their {\bf CDC} approach of (i) \underline{{\bf C}}ollating frequently anticipated questions and their authoritative answers, (ii) \underline{{\bf D}}istributing reliably by modeling multi-dimensional opinion networks with official information and discovering strategies to control them \cite{muppasani2024effectiveplanningstrategiesdynamic,infospread-planning-demo-Muppasani_Narayanan_Srivastava_Huhns_2024}, and (iii) helping people with diverse backgrounds \underline{{\bf C}}omprehend official information with personalization and provenance using chatbots \cite{safechat-elections-aimag,muppasani2025electionbot},  - all in the service of reducing information gap for increasing voter participation. But it all starts with authentic data.

In US, questions about state-specific election processes—such as voter registration, polling locations, absentee ballot rules, and early voting policies—are crucial for both voters and policymakers. However, the decentralized nature of U.S. elections means that this information is often fragmented across various state and local jurisdictions, creating barriers to accessibility and analysis.
AI has the potential to address these challenges by aggregating, standardizing, and analyzing election-related data.


Releasing datasets is a key tradition in advancing NLP research, often catalyzing further work in the field. Related datasets, such as Factify3M \cite{chakraborty-etal-2023-factify3m}, have enhanced online information reliability. Our dataset adheres to the NLP community's best practices.
\section{Resources and Methods} \label{sec:resources}

In this section, we describe the resources and methodologies utilized, including data, NLP techniques, analysis setup, and the novel \textbf{composite metrics} introduced for comparing US states.

\subsection{Data Preparation and Consolidation} \label{sec:data}





Election-related FAQs were compiled by extracting data from official election websites across all 50 U.S. states. State-specific data, stored as JSON files containing Q\&A pairs with metadata (state name, contributor, and timestamps), was consolidated into a unified dataset for analysis. Metadata was preserved for traceability and state-level topic analysis, while timestamps retained temporal context.  

Data preprocessing involved deduplication using \texttt{SequenceMatcher} \cite{difflib} (85\% similarity threshold) to eliminate semantic overlap, along with text cleaning to normalize formatting (e.g., whitespace, punctuation). Election-specific terms (e.g., URLs, ``voter-ID'') were preserved for domain relevance. The final dataset contained unique Q\&A pairs with metadata, optimized for topic modeling. Summary statistics, including total Q\&A counts, provided an analytical overview.


To analyze the U.S. 2024 election dataset, we processed question-answer pairs from official state election sources and a reputable non-profit. For all 50 states, we computed source-wise counts and statistical attributes, including the average, maximum, and minimum lengths (in alphanumeric characters) of questions and answers. Data, originally in JSON format, was cleaned to remove non-alphanumeric elements (e.g., escape sequences, hyperlinks) introduced during manual collection. These statistics, summarized in \cref{tab:stat_table}, guided dataset structuring and preprocessing for downstream analysis.

\subsection{Methods} \label{sec:method}

We use the following four standard NLP techniques to analyze the FAQs holistically. We conduct experiments for \textbf{Question ($Q$)}, \textbf{Answer ($A$)} and \textbf{Question + Answer ($Q$ + $A$)}. However, we mainly focus on $Q$ + $A$ while we also include additional results for only $Q$ and $A$ in the Appendix.

\subsubsection{Readability}

The U.S. 2024 election information provided by both the official state resources and the non-profit website is designed to help the public understand eligibility criteria, registration procedures, and the voting process in each state. Ensuring that this information is accessible to individuals of varying literacy levels is essential for assessing the quality of election resources. To evaluate this characteristic, we used the Python Textstat library \cite{Textstat} to perform a readability analysis on the dataset, including the questions, the answers, and complete pairs of questions and answers. The analysis used five standard readability metrics: Flesch-Kincaid Grade (FKG) \cite{FKG}, with scores ranging from 0-12 corresponding to US school grade levels; Gunning Fog Index (GFI) \cite{GFI}; SMOG Index (SI) \cite{SMOG}; Automated Readability Index (ARI) \cite{ARI}; and Coleman-Liau Index (CLI) \cite{CLI}. The latter four metrics range from 1–20+, with higher scores indicating more complex material, and scores above 13 on CLI suggesting content suitable for college-level readers and professionals.

\subsubsection{Summarization}


The quality of the U.S. election data from states and a non-profit organization relies on the alignment of answers to corresponding questions. Evaluating this alignment is challenging for lengthy state responses. To assess answer relevance, we summarized responses (350–800 characters) from all 50 states using extractive techniques via Python's Sumy library \cite{sumy} and abstractive methods with Hugging Face's DistilBART model \cite{DistilBART}. We evaluated summary quality using multiple metrics: ROUGE variants (ROUGE-1, ROUGE-2, ROUGE-L, ROUGE-W, ROUGE-S, ROUGE-SU) for unigram/bigram overlap, longest common subsequence, weighted n-gram overlap, skip-bigram overlap, and overall relevance. BLEU score measured n-gram overlap with reference questions, while cosine similarity assessed textual similarity to the original questions.


\subsubsection{Topic Analysis}

To perform the topic analysis, we utilized \textbf{Latent Dirichlet Allocation (LDA)}, a widely-used probabilistic model for identifying latent topics in text data. LDA is particularly effective for datasets like FAQs, where documents (in this case, question-answer pairs) can represent a mixture of multiple topics. 

To prepare the dataset for LDA, a document-term matrix (DTM) was constructed using TF-IDF (Term Frequency-Inverse Document Frequency) vectorization. This step involved transforming the text data into a numerical representation suitable for machine learning. Key preprocessing steps included:

\begin{itemize}[nolistsep]
    \item \textbf{Maximum Features:} The DTM was limited to the top 1000 most relevant terms to reduce noise while retaining informative features.
    \item \textbf{Stopword Removal:} Common English stopwords (e.g., ``the'', ``and'') were removed to focus on meaningful content.
    \item \textbf{n-gram Range:} Both unigrams and bigrams (e.g., ``voter registration'') were included to capture key phrases.
\end{itemize}

To determine the optimal number of topics, multiple topic counts ranging from 2 to 15 were evaluated using the following metrics:

\begin{itemize}[nolistsep]
    \item \textbf{Perplexity:} Measures the model's ability to generalize to unseen data, with lower values indicating better fit.
    \item \textbf{Silhouette Score:} Assesses the quality of document clustering within topics, with higher scores reflecting better-defined topics.
    \item \textbf{Topic Coherence:} Evaluates the semantic similarity of the top words in each topic, with higher scores indicating more interpretable topics.
\end{itemize}
Based on these metrics, the optimal number of topics was determined to be 8, balancing model complexity and interpretability.

The LDA model was then applied to the dataset with the following parameters:

\begin{itemize}[nolistsep]
    \item \textbf{Number of Topics:} 8
    \item \textbf{Maximum Iterations:} 20, ensuring convergence of the model.
    \item \textbf{Random State:} 42, for reproducibility of results.
\end{itemize}

The output of the LDA model included:
\begin{itemize}[nolistsep]
    \item \textbf{Topic-Word Distributions:} Highlighting the most representative words for each topic.
    \item \textbf{Document-Topic Distributions:} Indicating the proportion of each topic within each document.
\end{itemize}

These outputs provided interpretable and concise insights into the dominant themes within the dataset, forming the basis for further analysis and scoring.

\textbf{Topic Scoring}

To evaluate the quality and importance of the identified topics, a comprehensive scoring system was implemented. This system assigned equal weight (25\%) to four key components, ensuring a balanced assessment of each topic:

\begin{itemize}[nolistsep]
\item \textbf{Prevalence (25\%):} Measures the average proportion of a topic across all documents. Topics with higher prevalence are considered more central to the dataset, reflecting their widespread relevance.

\item \textbf{Coherence (25\%):} Evaluates the semantic consistency of the top words within each topic. Coherence was computed by analyzing the co-occurrence probabilities of the top 10 words within the same documents, ensuring the interpretability of the topic.

\item \textbf{Distinctiveness (25\%):} Measures how unique a topic is compared to others. This was calculated using Jensen-Shannon divergence, which quantifies the dissimilarity between topic distributions, emphasizing the uniqueness of high-scoring topics.

\item \textbf{Coverage (25\%):} Assesses the proportion of documents where a topic is significantly represented. A document was considered to ``cover'' a topic if its probability for that topic exceeded a threshold of 0.1. Topics with higher coverage scores were represented in a broader range of documents.

\end{itemize}

The final topic score for each topic was calculated as given in \cref{eq:final_topic_score}:

\begin{equation}
\scalebox{0.75}{$
\begin{aligned}
\text{Final Topic Score} = & \ 0.25 \times \text{Prevalence} 
+ \ 0.25 \times \text{Coherence} \\ 
+ & \ 0.25 \times \text{Distinctiveness} 
+ \ 0.25 \times \text{Coverage} 
\end{aligned}
$}
\label{eq:final_topic_score}
\end{equation}

Based on their final scores, topics were categorized into three priority levels:
\begin{itemize}[nolistsep]
    \item \textbf{High Priority:} Final score $>$ 0.7.
    \item \textbf{Moderate Priority:} Final score between 0.3 and 0.7.
    \item \textbf{Low Priority:} Final score $<$ 0.3.
\end{itemize}

This scoring system ensured that each topic was evaluated comprehensively, balancing its frequency, interpretability, uniqueness, and coverage within the dataset.

\subsubsection{Sentiment Analysis}

This study performed sentiment analysis on election-related data from various U.S. states, aiming to classify the sentiment of question-answer pairs as positive, negative, or neutral. The analysis used \textbf{VADER (Valence Aware Dictionary and sEntiment Reasoner)} \cite{vaderSentiment} to evaluate sentiment in short text segments, such as those in our dataset. VADER is effective for analyzing social media-like content and returns four sentiment scores: positive, neutral, negative, and a composite compound score, which ranges from -1 (extremely negative) to +1 (extremely positive), indicating the sentiment's direction and intensity. Each question-answer pair was assigned a sentiment based on its compound score:
\begin{itemize}[nolistsep]
        \item \textbf{Positive:} Compound score > 0.01
        \item \textbf{Negative:} Compound score < -0.01
        \item \textbf{Neutral:} Compound score between -0.01 and 0.01
    \end{itemize}

\subsection{Analyses setup} \label{sec:setup}

To get a holistic sense of all the metrics, we propose a novel score combing them. We propose a novel metric for FAQ Information Quality Score called \textbf{FIQS} (pronounced as \emph{``fix''}).

\paragraph{FIQS\textsubscript{voter}} incorporates sentiment analysis, readability assessment, and topic coverage evaluation. The underlying premise is that the voter prioritizes content comprehension while remaining indifferent to the mechanisms of its production (see \cref{eq:fiqs_voter}).

\vspace{-5mm}
\begin{equation}
\scalebox{0.8}{$
\begin{aligned}
\text{FIQS\textsubscript{voter}} = & \ 0.25 \times \text{Readability Score} \\
+ & \ 0.25 \times \text{Summarization Score} \\ 
+ & \ 0.25 \times \text{Sentiment Score} 
+ \ 0.25 \times \text{Topic Score} 
\end{aligned}
$}
\label{eq:fiqs_voter}
\end{equation}

\paragraph{FIQS\textsubscript{developer}} is evaluated based on sentiment, readability, topic coverage, and prompt relevance. The underlying premise is that the developer prioritizes not only comprehension but also the efficiency of content generation. Leveraging the pre-training capabilities of large language models (LLMs), we integrate them into the process to enhance efficiency (see \cref{eq:fiqs_dev}).

\vspace{-5mm}
\begin{equation}
\scalebox{0.8}{$
\begin{aligned}
\text{FIQS\textsubscript{developer}} = & \ 0.2 \times \text{Readability Score} \\
+ & \ 0.2 \times \text{Summarization Score} \\ 
+ & \ 0.2 \times \text{Sentiment Score} 
+ \ 0.2 \times \text{Topic Score} \\
+ & \ 0.2 \times \text{Prompt Relevance} 
\end{aligned}
$}
\label{eq:fiqs_dev}
\end{equation}
\section{Analyzing the state of FAQs} \label{sec:analysis}

We analyze and compare state-level data using individual and composite metrics, presenting results for \textbf{Question ($Q$)}, \textbf{Answer ($A$)} and \textbf{Question + Answer ($Q$ + $A$)}.

\begin{table}[!htp]\centering
\scriptsize
\begin{tabular}{lrrr}\toprule
& \textbf{mean} & \textbf{std. dev.} \\\midrule
\textbf{FIQS\textsubscript{voter}} &0.4084 &0.17 \\
\textbf{FIQS\textsubscript{developer}} &0.41832 &0.15 \\
\bottomrule
\end{tabular}
\caption{This table presents the mean and standard deviation for FIQS\textsubscript{voter} and FIQS\textsubscript{developer}.}
\label{tab:mean}
\end{table}

\begin{figure*}[ht]
\begin{subfigure}[b]{0.49\textwidth}
    \centering
    \includegraphics[width=\textwidth, height=5cm]{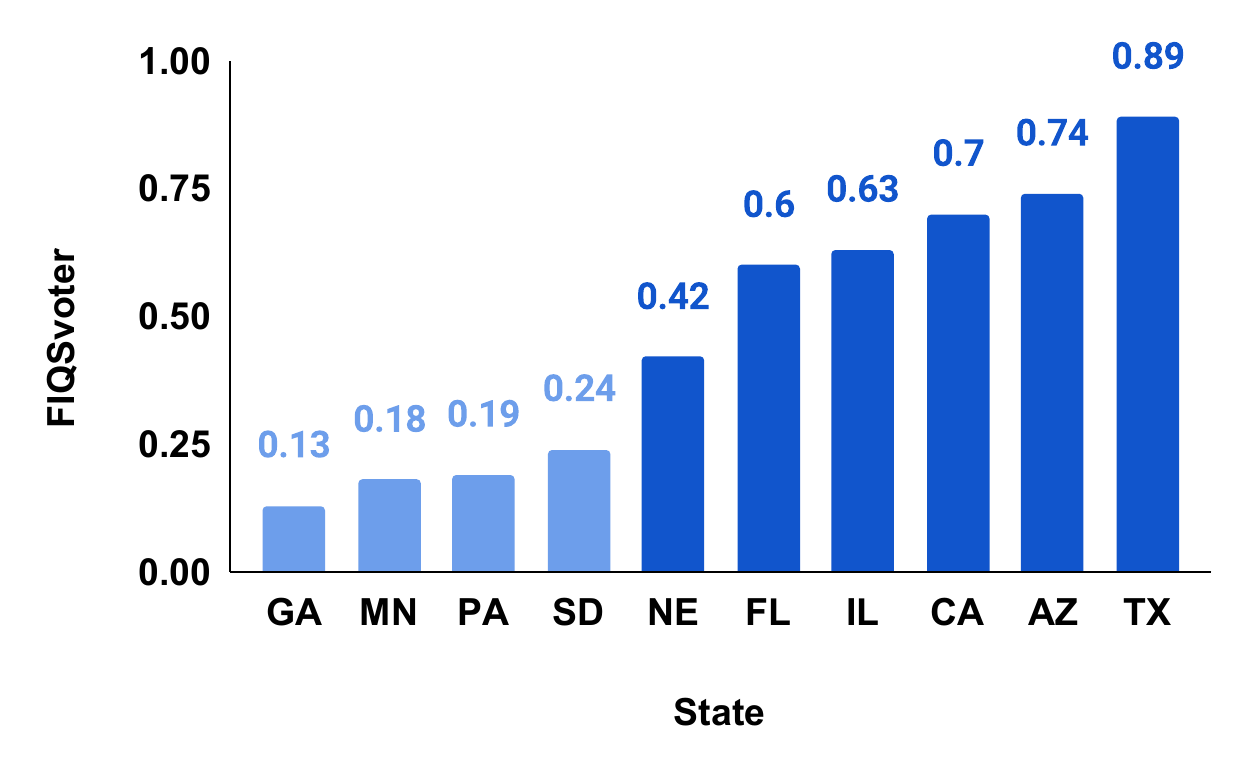}
    \label{fig:enter-label}
\end{subfigure}
\begin{subfigure}[b]{0.49\textwidth}
    \centering
    \includegraphics[width=\textwidth, height=5cm]{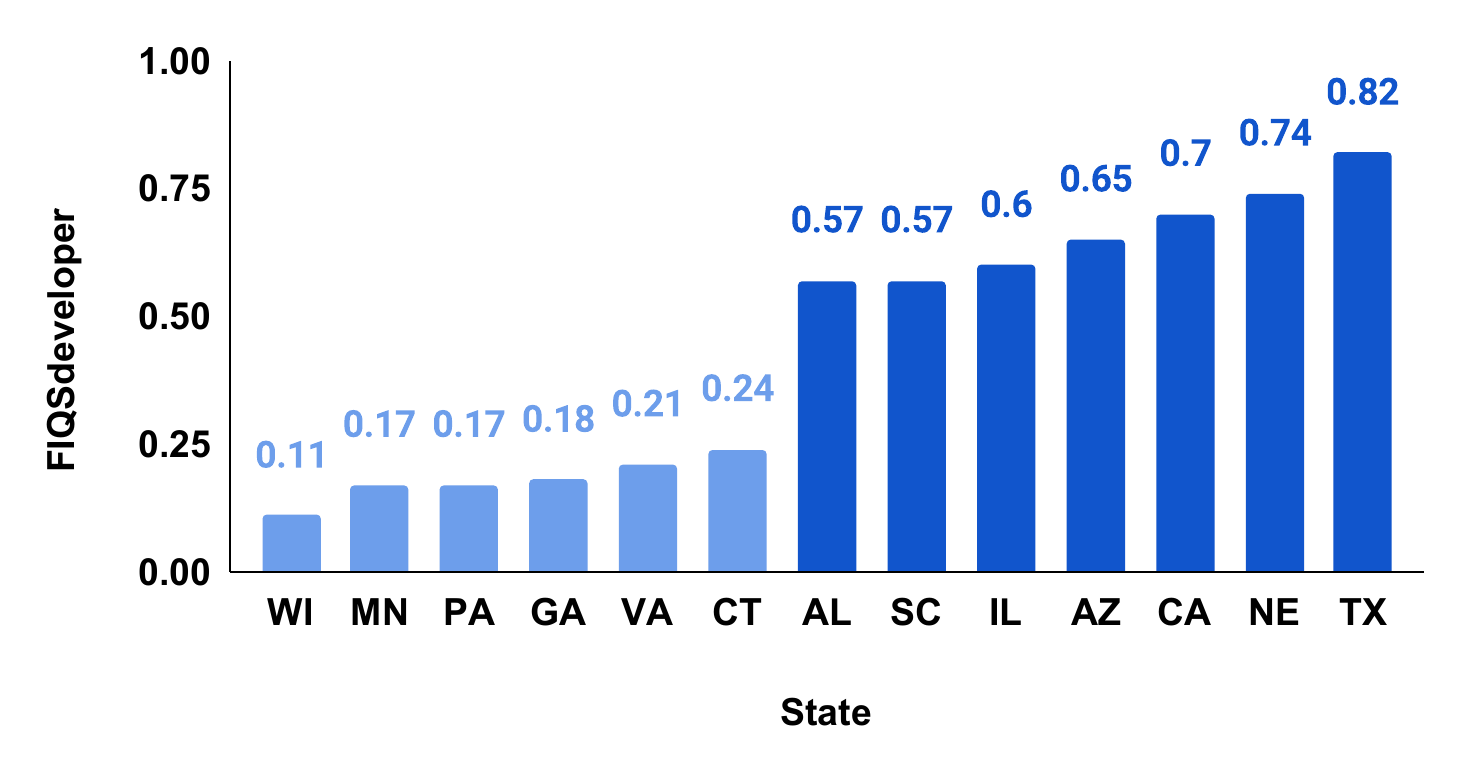}
    \label{fig:composite-comparison}
\end{subfigure}

\caption{US states leading and lagging in voter FAQ content quality, as assessed using cut-off of one standard deviation  from mean on the metric (i.e., $\geq$  ($\mu$ $\pm \sigma$); $\leq$ ($\mu$ $\pm \sigma$)).  We call them leaders and laggards, respectively.}
\label{fig:composite-comparison}
\end{figure*}

\begin{table*}[!ht]\centering
\caption{Question + Answer Topic Analysis Summary}\label{tab:qa_topic_analysis_sum }
\scriptsize
\resizebox{\textwidth}{!}{%
\begin{tabular}{lrrrrrrr}\toprule
\textbf{Topic} &\textbf{Final Score} &\textbf{Prevalence} &\textbf{Coherence} &\textbf{Distinctiveness} &\textbf{Coverage} &\textbf{Top Terms} \\\midrule
Topic 1: Political Parties and Primary Elections &461 &288 &443 &804 &308 &party, primary, political, election, primary election \\
Topic 2: Voter Registration &764 &826 &915 &521 &793 & registration, voter, voter registration, address, register \\
Topic 3: Absentee Voting &411 &499 &529 &88 &527 &ballot, absentee, absentee ballot, mail, return \\
Topic 4: Voting Equipment Security &163 &0 &0 &653 &0 &machines, write, secure, paper, card \\
Topic 5: Voter Identification Requirements &239 &66 &58 &694 &137 &photo, id, photo id, business, report \\
Topic 6: Military and Overseas Voting &392 &157 &0.19 &1 &223 &overseas, military, vote, register, register vote \\
Topic 7: Campaign Filing and Candidates &259 &65 &86 &771 &114 &campaign, candidates, filing, committee \\
Topic 8: Election Day and Polling Information &0.75 &1 &1 &0 &1 &election, ballot, day, voting, polling \\
\bottomrule
\end{tabular}%
}
\end{table*}

\subsection{Readability}

Our results revealed that questions in the dataset consistently received lower readability scores, indicating they were easier to comprehend. Conversely, the answers scored higher, suggesting greater complexity, likely due to the inclusion of specialized vocabulary and a focus on precision over simplicity. To identify the states with the highest ease of readability, all readability metrics were averaged and ranked by their lowest average scores. The readability scores for the question, answer, and combined question and answer are presented in \cref{tab:q_readability_table}, \cref{tab:a_readability_table}, and \cref{tab:qa_readability_table}, respectively. Please refer to \cref{app:read} for more details. 

The top three states in this category were Georgia, Maryland, and Pennsylvania. In contrast, states such as North Carolina, California, and Louisiana presented election information at higher levels of reading complexity.

\subsection{Summarization} 
To identify the states that provide the highest quality answers, we averaged the relevance metrics for each state. The states achieving the highest average relevance scores were deemed the leaders in response quality. From our analysis, Delaware, Kansas, and Michigan emerged as the top three performers, indicating superior alignment between their answers and corresponding questions. In contrast, Massachusetts, Rhode Island, and Hawaii scored the lowest, suggesting room for improvement in the quality of their responses. The summarization analysis for the question is presented for both \textit{Abstractive} (\cref{tab:a_q_sum_analysis}) and \textit{Extractive} (\cref{tab:e_q_sum_analysis}) approaches, while the analysis for the answer is provided for \textit{Abstractive} (\cref{tab:a_a_sum_analysis}) and \textit{Extractive} (\cref{tab:e_a_sum_analysis}) methods. Additionally, the summarization analysis for the combined question and answer is shown for \textit{Abstractive} (\cref{tab:a_qa_sum_analysis}) and \textit{Extractive} (\cref{tab:e_qa_sum_analysis}). Please refer to \cref{app:summ} for more details.

\subsection{Topic Analysis}

The topic modeling and scoring system results were analyzed to identify key topics in the election-related FAQ dataset. Each topic was evaluated based on its final score, component contributions (Prevalence, Coherence, Distinctiveness, and Coverage), and its distribution across states. Various visualization techniques were used to present these findings. Below are the key metrics:

\begin{itemize}[nolistsep]
    \item \textbf{Top Words:} Most representative words based on their probabilities in the topic-word distribution.
    \item \textbf{Prevalence:} Average proportion of the topic across all documents.
    \item \textbf{Coherence:} Semantic consistency of the top words.
    \item \textbf{Distinctiveness:} Uniqueness of the topic relative to others.
    \item \textbf{Coverage:} Proportion of documents where the topic probability exceeds 0.1.
\end{itemize}

A heatmap was generated to visualize the distribution of topics across states \cref{fig:topic_distribution_heatmap}.

The state performance in election FAQ coverage is assessed using a comprehensive scoring formula that incorporates multiple components (see \cref{eq:final_state_score}).
\vspace{-5mm}
\begin{equation}
\scalebox{0.8}{$
\begin{aligned}
\text{Final State Score} = 
& \textstyle \sum 
( \text{Topic\_Value} 
 \times \text{Topic\_Weight} \\
& \times \text{FAQ\_Normalization}
 \times \text{FAQ\_Penalty} ) 
\end{aligned}
$}
\label{eq:final_state_score}
\end{equation}

where, \textit{Topic\_Value} represents the state's coverage of each topic based on topic distribution, \textit{Topic\_Weight} indicates each topic's importance from the final analysis scores, and \textit{FAQ\_Normalization} and \textit{FAQ\_Penalty} are defined in \cref{eq:faq_normalize,eq:faq_penalty}.
\vspace{-5mm}

\begin{equation}
\scalebox{0.75}{$
\begin{aligned}
\text{FAQ\_Normalization} &= 
\frac{\text{state\_faq\_count}}{\text{max\_faq\_count}}
\end{aligned}
$}
\label{eq:faq_normalize}
\end{equation}

\begin{equation}
\scalebox{0.75}{$
\begin{aligned}
\text{FAQ\_Penalty} &= \min\left(1.0, \frac{\text{faq\_count}}{20}\right)
\end{aligned}
$}
\label{eq:faq_penalty}
\end{equation}

\cref{tab:qa_topic_analysis_sum } presents the distribution of final topic scores across the eight identified topics.
The topic analysis for the question, answer, and combined question and answer is provided in \cref{tab:question_topic_analysis}, \cref{tab:answer_topic_analysis}, and \cref{tab:qa_topic_analysis}, respectively. Please refer to \cref{app:topic} for more details.

The analysis revealed significant variations in
state performance. Michigan emerged as the leading state with a score of 0.572, supported by a comprehensive collection of 123 FAQs and strong coverage across all topics, particularly in Administrative \& Filing (0.271) and Voter Registration (0.141). Florida followed with a score of 0.413 and 111 FAQs, demonstrating well-balanced coverage across topics. Nevada (0.389), Oklahoma (0.388), and North Carolina (0.344) completed the top five, each maintaining robust FAQ counts above 75 and showing strong performance in key topics like Voter Registration and Administrative procedures.
Conversely, the analysis identified states with significant room for improvement. Mississippi ranked lowest with a score of 0.002, primarily due to having the least FAQs, resulting in minimal coverage across all topics. Similar patterns emerged for South Dakota (0.013), Wisconsin (0.015), Nebraska (0.019), and Montana (0.029), all characterized by FAQ counts below 12 and consequently limited topic coverage.

\subsection{Sentiment Analysis}

To identify the leaders and laggards in sentiment analysis across U.S. states, we analyzed the average compound sentiment scores obtained from VADER. The compound score, was used as the primary metric to assess the sentiment polarity and intensity associated with election-related FAQs. States with the highest average compound scores were identified as leaders, while those with the lowest average compound scores were categorized as laggards. 
 
\noindent \textbf{Leaders:} These states exhibited a higher proportion of positive sentiment, emphasizing optimistic and clear communication in their FAQs.\\
\textbf{Top 3 Leaders:} were Nebraska (0.380), Texas (0.372) and Arizona (0.327); average score in ().


\noindent \textbf{Laggards:} These states displayed a higher proportion of negative sentiment, potentially due to the phrasing of FAQs, lack of clarity, or underlying concerns in the election-related context. \\
\textbf{Top 3 Laggards:} were South Dakota (-0.053), Alaska (-0.068)
    and Wisconsin (-0.097); noting average score in ().

The detailed results for \textbf{Question} (\cref{tab:q_sentiment_analysis})
\textbf{Answer} (\cref{tab:a_sentiment_analysis})
\textbf{Question + Answer} (\cref{tab:qa_sentiment_analysis}) are given in \cref{app:sent}.

\subsection{Analyzing Questions for State Specificity}

Ensuring accessibility and informed decision-making requires election information provided by US state authorities and non-profit organizations to maintain a balance between generic and specific questions. Generic questions facilitate accessibility for voters with limited prior knowledge, such as first-time voters, by addressing fundamental aspects of the voting process. In contrast, specific questions localize information to the unique procedures and requirements of each state, enabling more precise voter guidance. This study conducted a specificity analysis on the questions from QA pairs across all 50 states to assess the balance between generic and specific content. A key aim was to identify commonalities in language across the questions posed by different states. If a state’s questions were similar to those of other states, they were classified as generic. For instance, a typical question such as ``Who can register to vote?'' is found in some form in many states' QA datasets, making it generic. On the other hand, state-specific questions feature distinctive language relevant to that jurisdiction, such as ``How do I obtain a document to prove I’m registered to vote in Hawaii?''

Our methodology involved several steps. First, we extracted the set of questions from each state’s dataset and processed them by removing stop words using the Natural Language Toolkit (NLTK) library to isolate key terms and focus on substantive content. We then generated sentence embeddings for each question using the Sentence Transformer model, providing a numerical representation of the semantic content of the questions. To account for variations in the number of questions across states, we normalised the embeddings, ensuring fair comparisons. Finally, we measured the similarity between questions using pairwise cosine similarity. A similarity threshold of 0.8 was used to classify questions as generic, while pairs with a similarity score of 1.0 were excluded to account for potential duplicate questions within states. \cref{fig:specificity-analysis}  in the Appendix, illustrates our findings, plotting the number of generic versus specific questions for each state. This visualization highlights trends in how states balance these two types of content, offering insights into the consistency and localization of voter information across the United States.

\begin{figure*}[ht] 
    \centering
    \begin{minipage}{0.6\textwidth} 
        \includegraphics[width=\linewidth]{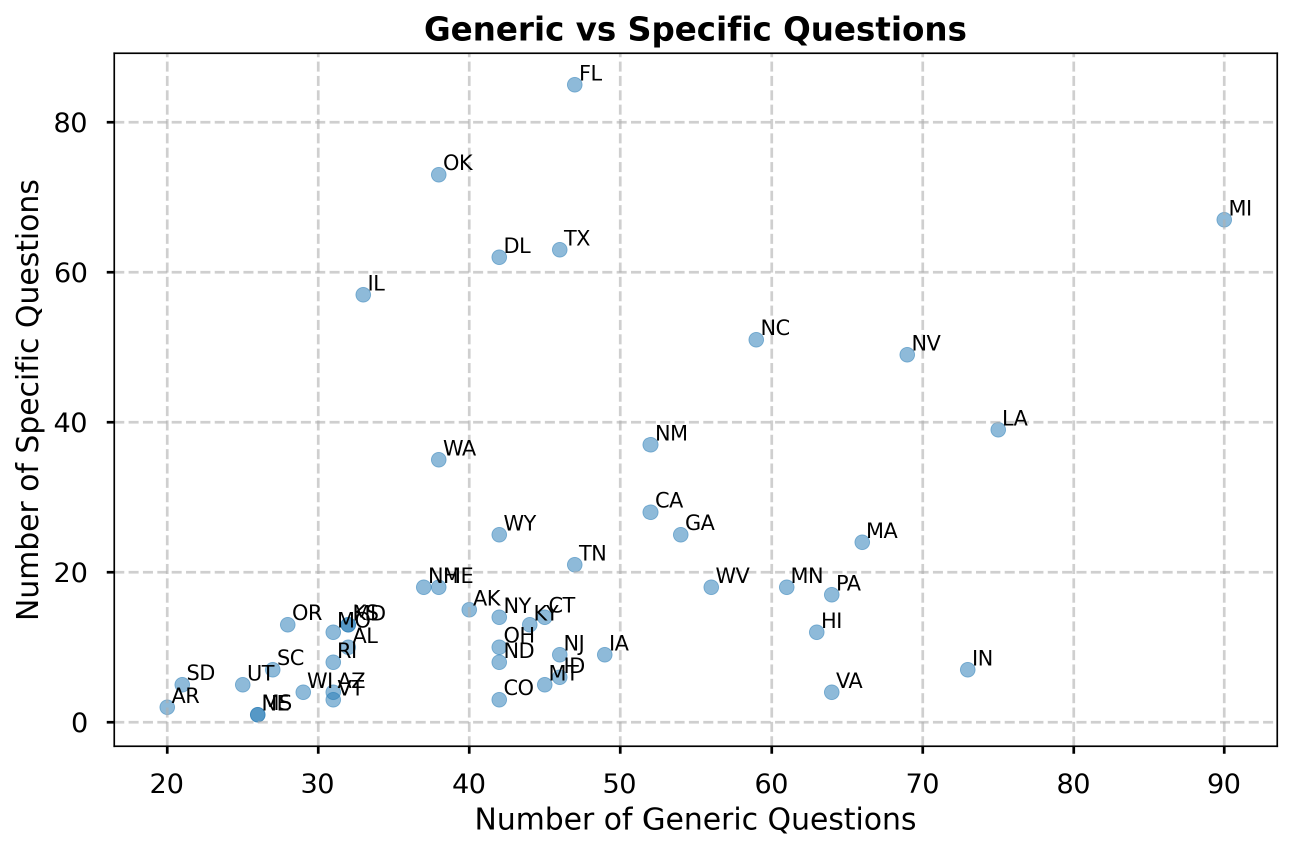} 
    \end{minipage}%
    \hfill
    \begin{minipage}{0.35\textwidth} 
        \caption{Scatter Plot of Generic vs Specific Questions Across States.
This scatter plot illustrates the distribution of generic and specific questions across the QA datasets of all 50 US states. Generic questions, which address fundamental aspects of the voting process, are plotted against specific questions, which localize information to state-specific procedures and requirements. The plot highlights the balance maintained by each state in providing voter information, with clusters indicating common trends and outliers suggesting unique patterns of question specificity.}
        \label{fig:specificity-analysis}
    \end{minipage}
\end{figure*}


\subsection{Prompt Analysis}
Since Large Language Models (LLMs) are being used in NLP tasks extensively, we also wanted to analyze how the FAQ content is amenable to LLM-based processing. 
In this study, we finetune a pretrained LLM specifically Llama-3.1-8B \cite{dubey2024llama} on the election dataset. The overall fine-tuning process involves the following steps:

\noindent \textbf{1. Dataset Preparation} 
The initial dataset for each state consists of question-answer pairs with metadata (source, timestamp, and state). For fine-tuning, the dataset is loaded via the Datasets library \cite{lhoest-etal-2021-datasets}, converted into a conversational format, and augmented with schema details in the system message. This enables fine-tuning with additional context. See \cref{fig:chat template} for the template.
 
\noindent \textbf{2. Model Fine-tuning} 
We fine-tuned the Meta LLaMA-3.1-8B model using the SFTTrainer from trl (Transformer Reinforcement Learning) \cite{vonwerra2022trl}, integrated with PEFT for efficient LLM tuning via QLoRA. The training used LoRA configurations with a learning rate of 2e-4, 3\% warmup, and a constant scheduler. The dataset was split 80\% for training, 10\% for validation, and 10\% for testing. Optimization employed AdamW with weight decay, adaptive learning rates, and cross-entropy loss for causal modeling. The model trained for 10 epochs with a batch size of 4, 2 gradient accumulation steps, and memory optimization via gradient checkpointing, 4-bit quantization, NF4, and bfloat16. Gradient clipping was applied with a max norm of 0.3, and LoRA had an alpha of 128, dropout of 0.05, and rank 256. Training took 11 hours on a Tesla V100-PCIE-32GB.

\noindent  \textbf{3. Evaluation}
The fine-tuned model is assessed on downstream tasks such as Readability, Summarization, Topic Modeling, and Sentiment Analysis, and evaluated by generating answers to training questions. Performance is measured using the same metrics as for the original question-answer pairs. 
We do not conduct experiments for questions, as the LLM solely generates responses without altering the input questions. Consequently, the outcomes remain consistent with previous results. The results are systematically presented in the formats of \textbf{Answer} and \textbf{Question + Answer}. Following this structure, readability results are detailed in (\cref{tab:llm_readability_analysis_a})
and (\cref{tab:llm_readability_analysis_qa}). Similarly, both \textit{Abstractive} and \textit{Extractive} summarization outcomes are organized in tables \cref{tab:llm_a_summary_analysis_a} and \cref{tab:llm_e_summary_analysis_a} for the answer, and \cref{tab:llm_a_summary_analysis_qa} and \cref{tab:llm_e_summary_analysis_qa} for the combined question and answer. Furthermore, topic analysis findings are provided in \cref{tab:llm_topic_analysis_a} and \cref{tab:llm_topic_analysis_qa}. Lastly, sentiment analysis results are also presented in \cref{tab:llm_sentiment_analysis_a} and \cref{tab:llm_sentiment_analysis_qa} (\cref{app:prompt}).

\section{Guidelines for improving the ecosystem} \label{sec:guidelines}

We note that 
Figure~\ref{fig:composite-comparison} gave a composite view of the leading and lagging US states in content quality, of which, 
 an illustration was shown in Figure~\ref{fig:intro-q}. Digging deeper, we found that leaders do a few things differently (and correctly) which others should follow. They are that leaders have more questions and answers (Table~\ref{tab:stat_table}) with  content that cover more topics (Table~\ref{tab: a_topic_analysis_sum}), that are readable (Tables~\ref{tab:q_readability_table},\ref{tab:a_readability_table},\ref{tab:qa_readability_table}),  and exhibit neutral sentiments (Figures~\ref{fig:senti-distrib},\ref{fig:senti-range}).

Based on these analyses, we provide the following guidelines for all states to improve their voter FAQ content. They are that states should (1) provide a reasonably large number of questions (typically $\geq$ 50)  covering a broad set of topics ($\geq$ five) in simple language,  (2) provide precise and specific answers which are  not too terse, (3) reduce overlap across questions by reducing overlap of topics, and (4) keep sentiment of content neutral. 
\section{Conclusion and Future Work} \label{sec:conclusion}

This paper makes many contributions starting by addressing the challenges faced by voters in finding answers to their election related questions by providing the {\bf first  dataset on Voter FAQs  covering all the
US states}.  Second, we introduce metrics for FAQ information quality score (FIQS)  with respect to  questions, answers, and answers to corresponding questions. Third, we use FIQS to analyze US FAQs to identify  leading, mainstream and lagging content practices and corresponding states. Finally, we identify what states across the spectrum can do to improve FAQ quality and thus,  the overall information ecosystem.

We provide verified, curated voter information to counteract widespread misinformation. This work, although promising, is just the first step. In future, one can work to remove the limitations and also build decision-support tools using the data to make effective tools available to voters. One can also separate the analysis by SECs, the primary, official data providers, and by secondary sources, e.g., non-profits like LWV, to understand how they complement each other.
\newpage
\section{Limitations}


Our work has a few limitations.
We have used open data of the 50 states as-is and relied on the unique position of SECs to provide authentic information about those states. Further, we have only used data from one non-profit, LWV, but it covers all the US states. The limitations can be easily overcome on number of sources can be overcome by adding more providers seamlessly. 
We have used state-of-the-art NLP methods as implemented in off-the-shelf tools; however, nothing precludes us from using new methods in future.

\section{Ethical Considerations} \label{sec:ethical}


We declare that all authors of this paper acknowledge the {\em ACM Code of Ethics} and honor the code of
conduct. This work collates and  evaluates {\bf open data} related to voters from SECs and the non-profit, LWV. By definition and spirit of open data, the data providers intended to make their data reusable and consumable, and we process it acknowledging full credits to providers. {\em (After anonymous review period, we will elaborate on our continued engagement with them.)} Our findings on content  is intended to help stakeholders understand best practices and help improve the overall voter information ecosystem. We believe our work will help the NLP community leverage its state-of-the-art methods to positively improve governance, an important aspect of society.

\bibliography{0_main}

\newpage
\appendix
\section{Appendix} \label{sec:appendix}

This section provides supplementary material in the form of additional examples, implementation details, etc. to bolster the reader's understanding of the concepts presented in this work.

\begin{figure*}[ht]
\begin{subfigure}[b]{0.49\textwidth}
    \centering
    \includegraphics[width=\textwidth]{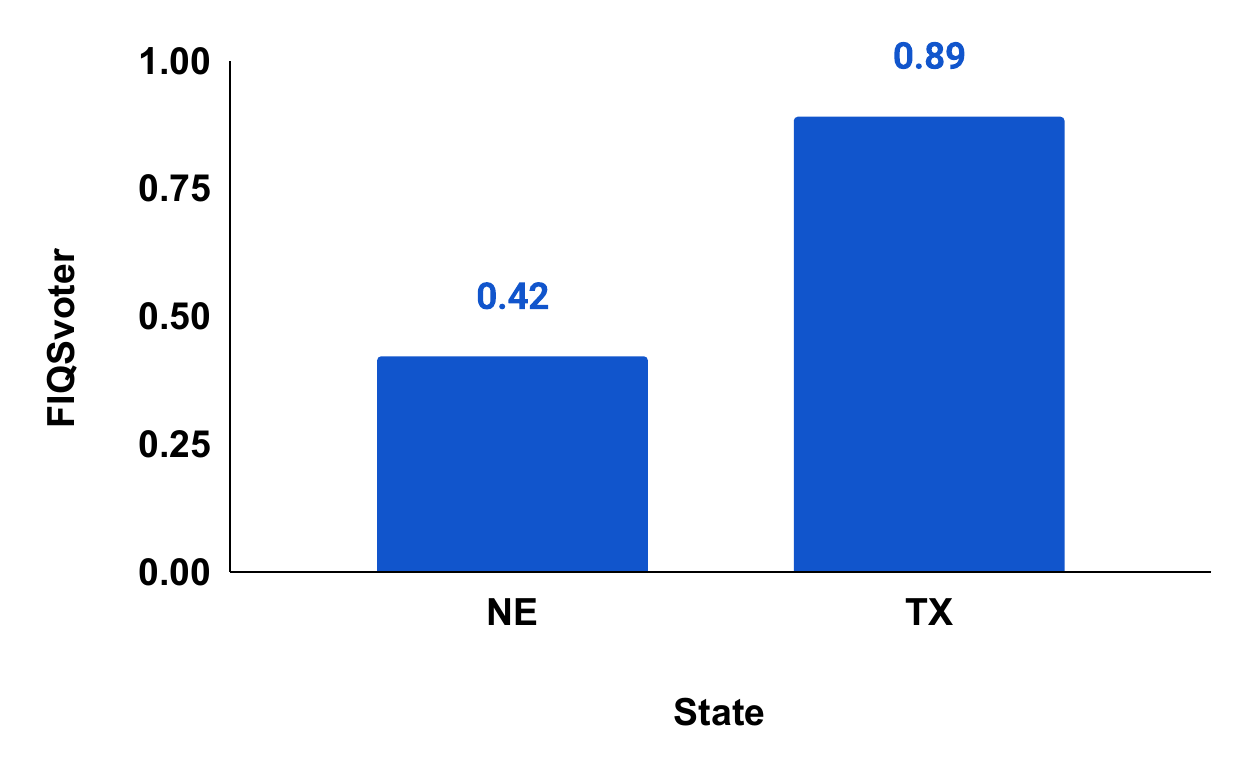}
    \label{fig:enter-label}
\end{subfigure}
\begin{subfigure}[b]{0.49\textwidth}
    \centering
    \includegraphics[width=\textwidth]{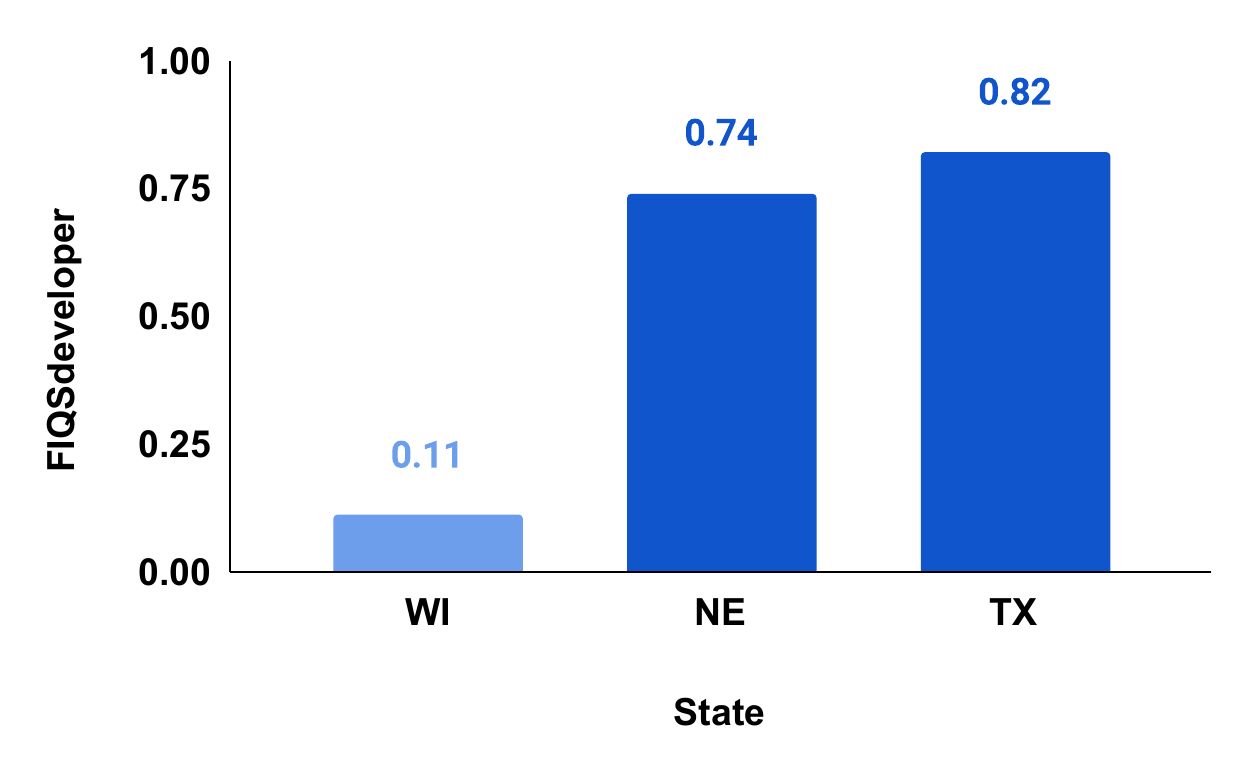}
    \label{fig:enter-label}
\end{subfigure}

\caption{US states leading and lagging in voter FAQ content quality, as assessed using cut-off of two standard deviation  from mean on the metric (i.e., $\geq$  ($\mu$ $\pm 2\sigma$); $\leq$ ($\mu$ $\pm 2\sigma$)).  
}
\label{fig:at-facts}
\end{figure*}

\section{Dataset} \label{app:data} 

Statistical analysis can reveal patterns, trends, and relationships within the data, providing insights into its underlying structure and significance with respect to our dataset. The results of our statistical analysis can be found in the (\cref{tab:stat_table}).

\section{Readability} \label{app:read}
Conducting readability assessments is important to ensure that the content within our dataset is clear, accessible, and effectively understood by the target audience. The readability scores for the question, answer, and the combined question and answer are provided in \cref{tab:q_readability_table}, \cref{tab:a_readability_table}, and \cref{tab:qa_readability_table}, respectively.

\section{Summarization} \label{app:summ}

The summarization ratio was dynamically calculated based on the token count of the original answers to maintain consistency between varying answer lengths. The code below illustrates how this value is computed:

\begin{verbatim}
    if answer_token_count < 200:
        reduction_ratio = 0.5
    elif answer_token_count < 300:
        reduction_ratio = 0.4
    else:
        reduction_ratio = 0.3
\end{verbatim}

The values 0.5, 0.4, and 0.3 correspond to reduction ratios of 50\%, 40\%, and 30\%, respectively. Specifically, if the answer is relatively short, with fewer than 200 tokens, the reduction ratio is set to 0.5, reflecting a moderate reduction. For answers of greater length, but fewer than 300 tokens, the reduction ratio is slightly reduced to 0.4. Finally, for answers comprising 300 tokens or more, the reduction ratio is further decreased to 0.3, signifying a smaller reduction for longer responses.

The summarization analysis for the question is provided for both the \textit{Abstractive} (\cref{tab:a_q_sum_analysis}) and \textit{Extractive} (\cref{tab:e_q_sum_analysis}) methods, while the analysis for the answer is presented for the \textit{Abstractive} (\cref{tab:a_a_sum_analysis}) and \textit{Extractive} (\cref{tab:e_a_sum_analysis}) approaches. Furthermore, the summarization analysis for the combined question and answer is displayed for both the \textit{Abstractive} (\cref{tab:a_qa_sum_analysis}) and \textit{Extractive} (\cref{tab:e_qa_sum_analysis}) techniques. Semantic Overlap is defined as a weighted average of cosine similarity and ROUGE-1 scores. While Abstractive summarization yields better results, it has the limitation of lacking provenance.

\section{Topic Analysis} \label{app:topic}
Topic analysis is crucial as it facilitates the identification and understanding of key themes and subject matter within our dataset, thereby enhancing content organization, relevance, and the ability to draw meaningful insights. The results of our topic analysis for the question, answer, and combined question and answer are provided in \cref{tab:question_topic_analysis}, \cref{tab:answer_topic_analysis}, and \cref{tab:qa_topic_analysis}, respectively.

\section{Sentiment Analysis} \label{app:sent}

\paragraph{Visualization:} To provide a clear understanding of the sentiment distribution across U.S. states, we visualized the data using a stacked bar chart sorted alphabetically by state names (Fig.2). Each bar represents the percentage distribution of positive, neutral, and negative sentiments for the election-related FAQs in that state. The comprehensive results for sentiment analysis are provided for questions (\cref{tab:q_sentiment_analysis}), answers (\cref{tab:a_sentiment_analysis}), and Question + Answer pairs (\cref{tab:qa_sentiment_analysis}).

\section{Prompt Analysis} \label{app:prompt}

\paragraph{LLama3.1} Large language model (LLMs) are a special class of pre-trained language model (PLMs). LLMs exhibit special capabilities due to their enormous size and pre-training on large amounts of text data, allowing them to achieve excellent performance in many natural language processing tasks without any task-specific training. Fine-tuning these LLMs involves adapting the pre-trained model to specific tasks. Specifically, the LLM is partially retrained using domain specific data. Llama 3.1 is an auto-regressive language model that uses an optimized transformer architecture. The model was pretrained on ~15 trillion tokens of data from publicly available sources.

\paragraph{Model Fine-tuning} The SFTTrainer makes it easy to supervise fine-tune  LLMs. The SFTTrainer is a subclass of the Trainer from the transformers library. It provides features such as Dataset formatting, including conversational and instruction format and PEFT (parameter-efficient fine-tuning) support including Q-LoRA. We use QLoRA to reduce the memory footprint of  the large language models during finetuning, without sacrificing performance by using 4-bit quantization. 

\textbf{Sentiment Analysis}

The sentiment of the predicted answers are compared to the sentiment of the original answer. Similar to the previous approach, VADER is used to label the sentiment of the original answer and the predicted answer.
From the Table it can be seen that the output of the model is mostly aligned with the actual answers in terms of sentiment. 

The results are systematically organized according to \textbf{Answer} and \textbf{Question + Answer}. In this structure, the readability outcomes are detailed in \cref{tab:llm_readability_analysis_a} and \cref{tab:llm_readability_analysis_qa}. Similarly, the results of both \textit{Abstractive} and \textit{Extractive} summarization are presented in \cref{tab:llm_a_summary_analysis_a} and \cref{tab:llm_e_summary_analysis_a} for the answer, and in \cref{tab:llm_a_summary_analysis_qa} and \cref{tab:llm_e_summary_analysis_qa} for the combined question and answer. Furthermore, the findings from the topic analysis are provided in \cref{tab:llm_topic_analysis_a} and \cref{tab:llm_topic_analysis_qa}, while the sentiment analysis results are discussed in \cref{tab:llm_sentiment_analysis_a} and \cref{tab:llm_sentiment_analysis_qa}.

\begin{figure*}[t]
\centering
\begin{tcolorbox}[colframe=brown, colback=yellow!10, coltitle=black, top=-2.5mm, bottom=-2.5mm, left=-0.5mm, right=-0.5mm]
\begin{lstlisting}[breaklines=true, basicstyle=\ttfamily\tiny]
{
  "messages": 
  [
    {"role": "system", "content": "You are an agent specialized in answering all questions related to the 2024 elections for various states in the United States. Users will ask you questions in English, and you will generate accurate and concise answers based on the specific state mentioned. State: Minnesota. Source: https://sos.state.mn.us/elections-voting/register-to-vote/common-registration-questions/#typo. Data Collection Timestamp: 2024-11-23 19:00:00"},
    {"role": "user", "content": "How do I fix a typo in my registration?"},
    {"role": "assistant", "content": "Contact your county election office."}
  ]
}
\end{lstlisting}
\end{tcolorbox}
\caption{Question Template Example}
\label{fig:chat template}
\end{figure*}

\clearpage
\begin{table*}
    \centering
    \scalebox{0.6}{%
    \begin{tabular}{l | c | c | c | c | c | c | c | c }
    \hline
    \hline
    States & Official QA Pairs & Non-Profit QA Pairs & Question Average & Question Longest & Question Shortest & Answer Average & Answer Longest & Answer Shortest \\
    \hline
    Alabama & 20 & 22 & 49.881 & 95 & 14 & 400.881 & 3579 & 25\\
    Alaska & 32 & 23 & 45.982 & 307 & 17 & 441.564 & 2771 & 30\\
    Arizona & 13 & 22 & 60.429 & 95 & 20 & 533.857 & 1702 & 30\\
    Arkansas & 4 & 18 & 36.364 & 57 & 20 & 358.727 & 960 & 28\\
    California & 57 & 23 & 60.025 & 178 & 10 & 508.899 & 3650 & 3\\
    Colorado & 22 & 23 & 41.044 & 105 & 17 & 579.178 & 2556 & 35\\
    Connecticut & 37 & 22 & 38.390 & 115 & 11 & 472.220 & 2023 & 44\\
    Delaware & 82 & 22 & 63.644 & 268 & 16 & 376.356 & 3470 & 2\\
    Florida & 110 & 22 & 54.705 & 196 & 14 & 390.008 & 3416 & 13\\
    Georgia & 50 & 29 & 43.051 & 135 & 12 & 247.101 & 2168 & 37\\
    Hawaii & 52 & 23 & 31.453 & 61 & 10 & 408.600 & 1942 & 28\\
    Idaho & 37 & 15 & 35.000 & 65 & 17 & 279.269 & 1227 & 22\\
    Illinois & 0 & 103 & 148.039 & 613 & 4 & 13.155 & 19 & 12\\
    Indiana & 58 & 22 & 40.288 & 86 & 10 & 413.712 & 2677 & 20\\
    Iowa & 36 & 22 & 47.448 & 96 & 18 & 395.052 & 2558 & 22\\
    Kansas & 24 & 21 & 50.044 & 194 & 26 & 441.644 & 2088 & 30\\
    Kentucky & 34 & 23 & 48.579 & 86 & 19 & 457.684 & 3723 & 42\\
    Louisiana & 93 & 21 & 35.404 & 72 & 12 & 789.842 & 5757 & 40\\
    Maine & 33 & 23 & 41.161 & 131 & 16 & 368.696 & 2374 & 62\\
    Maryland & 23 & 22 & 34.956 & 66 & 10 & 425.733 & 1547 & 35\\
    Massachusetts & 68 & 22 & 29.300 & 56 & 11 & 456.411 & 2521 & 4\\
    Michigan & 135 & 22 & 50.255 & 131 & 20 & 529.994 & 9549 & 44\\
    Minnesota & 23 & 22 & 34.956 & 66 & 10 & 425.733 & 1547 & 35\\
    Mississippi & 5 & 22 & 59.667 & 86 & 12 & 435.963 & 2966 & 23\\
    Missouri & 24 & 19 & 45.744 & 154 & 12 & 486.930 & 2145 & 0\\
    Montana & 13 & 37 & 39.260 & 90 & 20 & 271.520 & 1029 & 35\\
    Nebraska & 10 & 17 & 36.222 & 55 & 12 & 481.148 & 1629 & 68\\
    Nevada & 79 & 39 & 45.593 & 120 & 13 & 304.720 & 2092 & 3\\
    New Hampshire & 34 & 21 & 42.909 & 106 & 17 & 451.945 & 2021 & 2 \\
    New Jersey & 33 & 22 & 57.109 & 171 & 20 & 379.200 & 2337 & 22\\
    New Mexico & 66 & 23 & 56.250 & 126 & 18 & 354.648 & 3107 & 22\\
    New York & 20 & 36 & 43.804 & 110 & 25 & 415.482 & 1885 & 33\\
    North Carolina & 88 & 22 & 53.782 & 139 & 17 & 463.464 & 3841 & 3\\
    North Dakota & 28 & 22 & 30.980 & 61 & 10 & 534.080 & 3760 & 37\\
    Ohio & 30 & 22 & 49.942 & 86 & 17 & 430.692 & 3759 & 26\\
    Oklahoma & 89 & 22 & 62.252 & 144 & 12 & 322.045 & 2612 & 23\\
    Oregon & 19 & 22 & 41.537 & 83 & 15 & 439.585 & 1801 & 25\\
    Pennsylvania & 58 & 23 & 51.716 & 142 & 10 & 299.716 & 2315 & 8\\
    Rhode Island & 17 & 22 & 34.308 & 89 & 17 & 492.513 & 2309 & 44\\
    South Carolina & 23 & 11 & 55.824 & 115 & 18 & 395.235 & 1460 & 21\\
    South Dakota & 8 & 18 & 44.269 & 98 & 17 & 412.731 & 1484 & 25\\
    Tennessee & 50 & 18 & 43.029 & 115 & 12 & 309.132 & 1581 & 37\\
    Texas & 69 & 40 & 62.844 & 213 & 16 & 604.101 & 5777 & 28\\
    Utah & 13 & 17 & 42.633 & 78 & 10 & 319.867 & 1132 & 25\\
    Vermont & 13 & 21 & 37.059 & 92 & 16 & 354.059 & 1860 & 35\\
    Virginia & 19 & 49 & 38.162 & 78 & 20 & 381.676 & 1715 & 22\\
    Washington & 51 & 22 & 33.329 & 79 & 13 & 421.507 & 1694 & 44\\
    West Virginia & 52 & 22 & 51.284 & 98 & 19 & 318.041 & 2767 & 35\\
    Wisconsin & 4 & 29 & 39.970 & 93 & 20 & 496.424 & 3021 & 24\\
    Wyoming & 49 & 18 & 44.403 & 1222 & 12 & 336.448 & 1628 & 41\\
    \hline
    \hline
\end{tabular}%
}
    \caption{This table presents a detailed analysis of question-answer (QA) pairs for each U.S. state. It includes the number of QA pairs sourced from the official state voting websites (Official QA Pairs) and from non-profit voting websites (Non-Profit QA Pairs). Additionally, it provides the average question length in alphanumeric characters (Question Average), the length of the longest question (Question Longest), and the length of the shortest question (Question Shortest). Similarly, for answers, it lists the average length (Answer Average), the length of the longest answer (Answer Longest), and the length of the shortest answer (Answer Shortest).}
    \label{tab:stat_table}
\end{table*}

\begin{table*}
    \centering
    \scalebox{0.8}{%
\begin{tabular}{l | c | c | c | c | c }
        \hline
        \hline
        States & FKG & GFI & SI & ARI & CLI\\
        \hline
        Alabama & 5.748 & 8.414 & 0.338 & 5.502 & 7.393\\
        Alaska & 6.473 & 9.359 & 0.427 & 5.889 & 7.512\\
        Arizona & 7.483 & 10.238 & 0.000 & 7.229 & 8.677\\
        Arkansas & 5.450 & 8.174 & 0.000 & 4.964 & 7.084\\
        California & 8.104 & 11.087 & 0.000 & 8.194 & 9.833\\
        Colorado & 5.122 & 7.363 & 0.000 & 4.820 & 6.438\\
        Connecticut & 4.905 & 8.075 & 0.000 & 3.481 & 4.851\\
\hline
        \hline
    \end{tabular}%
    }
    \caption{The readability analysis scores for the questions from each state.}
    \label{tab:q_readability_table}
\end{table*}

\begin{table*}
    \centering
    \scalebox{0.8}{%
    \begin{tabular}{l | c | c | c | c | c }
        \hline
        \hline
        States & FKG & GFI & SI & ARI & CLI\\
        \hline
        Alabama & 9.936 & 11.281 & 7.405 & 11.536 & 10.741\\
        Alaska & 10.455 & 11.690 & 6.951 & 12.065 & 11.106\\
        Arizona & 9.983 & 11.026 & 8.329 & 11.671 & 10.981\\
        Arkansas & 12.773 & 14.875 & 6.005 & 14.568 & 10.065\\
        California & 12.281 & 13.158 & 8.515 & 14.247 & 12.010\\
        Colorado & 10.569 & 11.411 & 8.162 & 12.678 & 11.819\\
        Connecticut & 10.842 & 12.504 & 8.441 & 12.520 & 10.681\\
\hline
        \hline
    \end{tabular}%
    }
    \caption{The readability analysis score for the answers from each state.}
    \label{tab:a_readability_table}
\end{table*}

\begin{table*}
    \centering
    \begin{tabular}{l | c | c | c | c | c }
				\hline
				\hline
				States & FKG & GFI & SI & ARI & CLI\\
				\hline
				Alabama & 8.819 & 9.707 & 9.264 & 9.767 & 9.927\\
				Alaska & 9.424 & 10.345 & 9.260 & 10.453 & 10.568\\
				Arizona & 9.569 & 10.203 & 9.966 & 10.809 & 10.633\\
				Arkansas & 10.127 & 11.910 & 9.105 & 11.227 & 9.559\\
				California & 11.256 & 11.699 & 11.303 & 12.725 & 11.978\\
				Colorado & 9.487 & 9.964 & 9.749 & 11.096 & 10.972\\
				Connecticut & 9.349 & 10.600 & 10.837 & 10.297 & 9.866\\
\hline
				\hline
			\end{tabular}
    \caption{The readability analysis score for each state's question and answer pairs.}
    \label{tab:qa_readability_table}
\end{table*}

\begin{table*}\centering
\caption{Abstractive Question Summarization Analysis }\label{tab:a_q_sum_analysis}
\scriptsize
\begin{tabular}{lrrrrrrrrr}\toprule
State &Cosine Similarity &ROUGE-1 &ROUGE-2 &ROUGE-L &ROUGE-W &ROUGE-S &ROUGE-SU &BLEU Score \\\midrule
AK &0.13 &0.14 &0.03 &0.10 &0.06 &0.01 &0.02 &0.01 \\
AL &0.22 &0.22 &0.09 &0.17 &0.12 &0.02 &0.05 &0.02 \\
AR &0.20 &0.18 &0.06 &0.15 &0.12 &0.02 &0.04 &0.01 \\
AZ &0.18 &0.21 &0.06 &0.16 &0.12 &0.02 &0.05 &0.01 \\
CA &0.19 &0.19 &0.07 &0.14 &0.11 &0.03 &0.05 &0.02 \\
CO &0.17 &0.16 &0.05 &0.12 &0.08 &0.01 &0.03 &0.01 \\
CT &0.13 &0.14 &0.04 &0.11 &0.08 &0.01 &0.03 &0.01 \\
\bottomrule
\end{tabular}
\end{table*}

\begin{table*}\centering
\caption{Extractive Question Summary Analysis}\label{tab:e_q_sum_analysis}
\scriptsize
\begin{tabular}{lrrrrrrrrr}\toprule
State &Cosine Similarity &ROUGE-1 &ROUGE-2 &ROUGE-L &ROUGE-W &ROUGE-S &ROUGE-SU &BLEU Score \\\midrule
AK &0.031 &0.057 &0.005 &0.053 &0.021 &0.001 &0.005 &0.002 \\
AL &0.042 &0.037 &0.011 &0.031 &0.030 &0.005 &0.011 &0.004 \\
AR &0.032 &0.042 &0.000 &0.031 &0.030 &0.000 &0.007 &0.003 \\
AZ &0.060 &0.077 &0.023 &0.057 &0.040 &0.003 &0.014 &0.005 \\
CA &0.043 &0.064 &0.018 &0.050 &0.038 &0.007 &0.015 &0.005 \\
CO &0.063 &0.069 &0.019 &0.053 &0.035 &0.003 &0.011 &0.003 \\
CT &0.028 &0.044 &0.006 &0.035 &0.021 &0.000 &0.006 &0.002 \\
\bottomrule
\end{tabular}
\end{table*}

\begin{table*}\centering
\caption{Abstractive Answer Summary Analysis}\label{tab:a_a_sum_analysis}
\scriptsize
\begin{tabular}{lrrrrrrrrr}\toprule
State &Cosine Similarity &ROUGE-1 &ROUGE-2 &ROUGE-L &ROUGE-W &ROUGE-S &ROUGE-SU &BLEU Score \\\midrule
AK &0.712 &0.520 &0.493 &0.488 &0.425 &0.435 &0.459 &0.180 \\
AL &0.717 &0.515 &0.486 &0.481 &0.413 &0.422 &0.446 &0.173 \\
AR &0.743 &0.582 &0.557 &0.547 &0.477 &0.466 &0.492 &0.223 \\
AZ &0.718 &0.504 &0.484 &0.476 &0.422 &0.453 &0.475 &0.178 \\
CA &0.733 &0.550 &0.526 &0.528 &0.458 &0.464 &0.488 &0.203 \\
CO &0.719 &0.502 &0.477 &0.476 &0.409 &0.414 &0.438 &0.159 \\
CT &0.732 &0.561 &0.539 &0.541 &0.481 &0.486 &0.505 &0.222 \\
\bottomrule
\end{tabular}
\end{table*}

\begin{table*}\centering
\caption{Extractive Answer Summary Analysis}\label{tab:e_a_sum_analysis}
\scriptsize
\begin{tabular}{lrrrrrrrrr}\toprule
State &Cosine Similarity &ROUGE-1 &ROUGE-2 &ROUGE-L &ROUGE-W &ROUGE-S &ROUGE-SU &BLEU Score \\\midrule
AK &0.167 &0.068 &0.065 &0.068 &0.058 &0.065 &0.072 &0.000 \\
AL &0.185 &0.093 &0.092 &0.093 &0.083 &0.096 &0.100 &0.007 \\
AR &0.125 &0.058 &0.056 &0.058 &0.052 &0.059 &0.063 &0.007 \\
AZ &0.258 &0.135 &0.132 &0.135 &0.118 &0.137 &0.145 &0.012 \\
CA &0.177 &0.081 &0.079 &0.081 &0.073 &0.083 &0.088 &0.007 \\
CO &0.273 &0.126 &0.123 &0.126 &0.112 &0.127 &0.134 &0.015 \\
CT &0.140 &0.067 &0.065 &0.067 &0.060 &0.065 &0.069 &0.017 \\
\bottomrule
\end{tabular}
\end{table*}

\begin{table*}[!htp]\centering
\caption{Question + Answer Abstractive Summary Analysis}\label{tab:a_qa_sum_analysis}
\scriptsize
\begin{tabular}{lrrrrrrrrr}\toprule
State &Cosine Similarity &ROUGE-1 &ROUGE-2 &ROUGE-L &ROUGE-W &ROUGE-S &ROUGE-SU &BLEU Score \\\midrule
AK &0.727 &0.537 &0.512 &0.505 &0.438 &0.446 &0.470 &0.192 \\
AL &0.743 &0.536 &0.508 &0.521 &0.457 &0.455 &0.477 &0.189 \\
AR &0.782 &0.575 &0.549 &0.530 &0.460 &0.474 &0.499 &0.227 \\
AZ &0.707 &0.503 &0.482 &0.480 &0.424 &0.445 &0.467 &0.171 \\
CA &0.741 &0.533 &0.506 &0.519 &0.444 &0.433 &0.459 &0.181 \\
CO &0.724 &0.492 &0.467 &0.478 &0.413 &0.415 &0.436 &0.154 \\
CT &0.721 &0.544 &0.518 &0.526 &0.464 &0.461 &0.481 &0.202 \\
\bottomrule
\end{tabular}
\end{table*}

\begin{table*}[!htp]\centering
\caption{Question + Answer Extractive Summary Analysis}\label{tab:e_qa_sum_analysis}
\scriptsize
\begin{tabular}{lrrrrrrrrr}\toprule
State &Cosine Similarity &ROUGE-1 &ROUGE-2 &ROUGE-L &ROUGE-W &ROUGE-S &ROUGE-SU &BLEU Score \\\midrule
AK &0.151 &0.060 &0.057 &0.060 &0.051 &0.057 &0.062 &0.000 \\
AL &0.151 &0.077 &0.076 &0.077 &0.069 &0.079 &0.083 &0.005 \\
AR &0.201 &0.127 &0.125 &0.127 &0.120 &0.128 &0.132 &0.071 \\
AZ &0.219 &0.101 &0.099 &0.101 &0.090 &0.103 &0.110 &0.007 \\
CA &0.203 &0.101 &0.098 &0.101 &0.091 &0.102 &0.109 &0.012 \\
CO &0.292 &0.144 &0.141 &0.144 &0.130 &0.147 &0.154 &0.030 \\
CT &0.123 &0.056 &0.055 &0.056 &0.050 &0.055 &0.058 &0.012 \\
\bottomrule
\end{tabular}
\end{table*}

\begin{figure*}
    \centering
    \includegraphics[width=\linewidth]{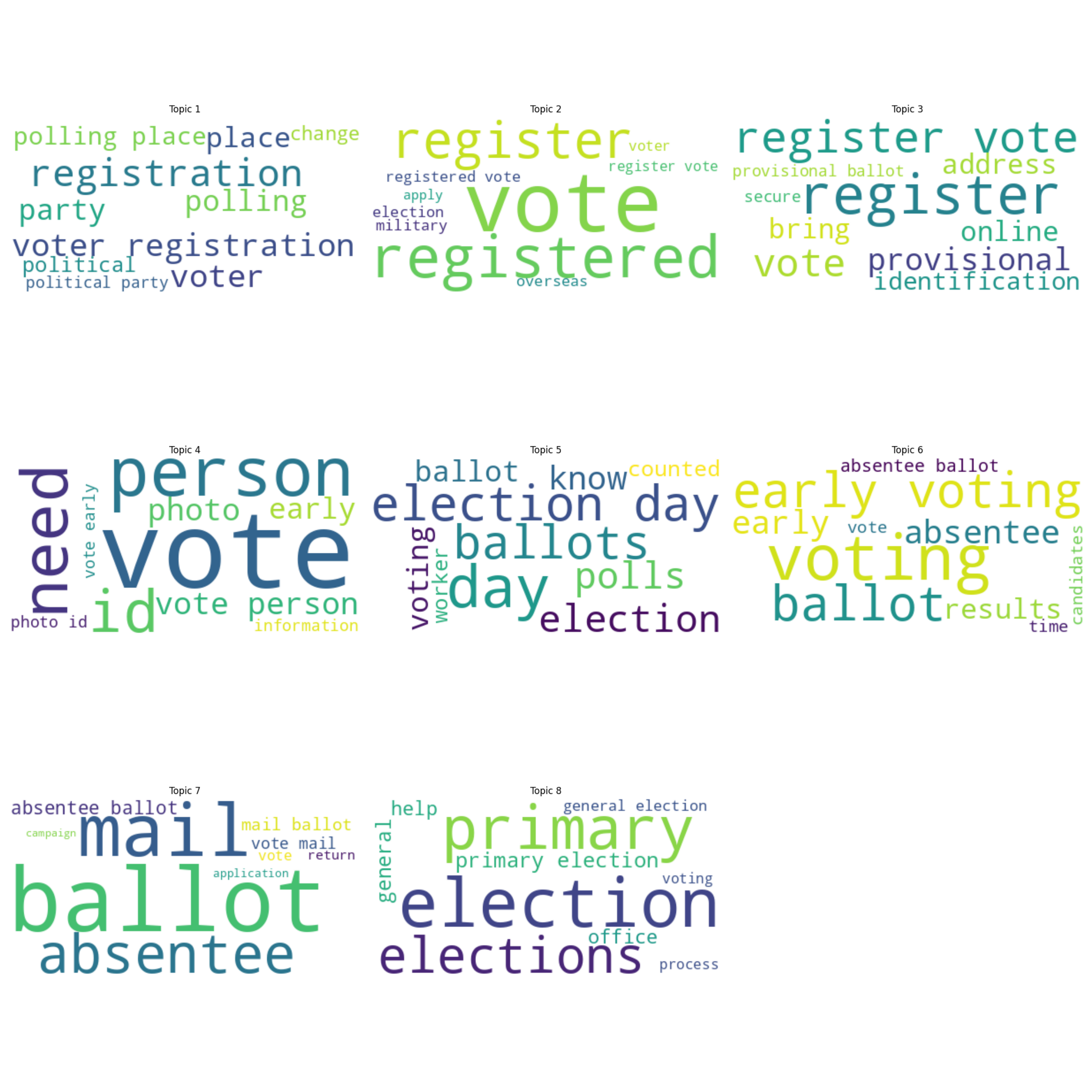}
    \caption{Question Topic Analysis Word Tag Cloud}
    \label{fig:enter-label}
\end{figure*}

\begin{figure*}
    \centering
    \includegraphics[width=\linewidth]{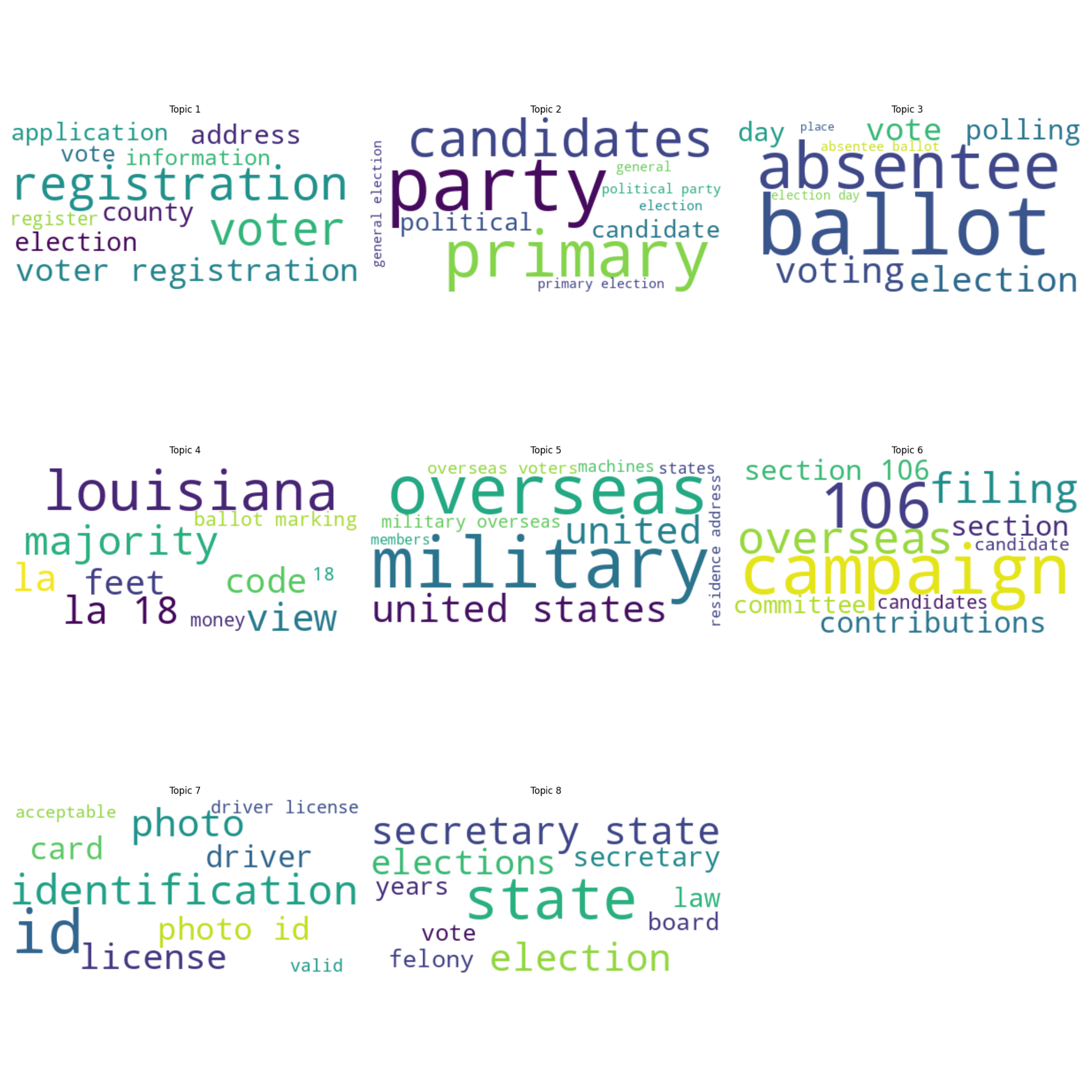}
    \caption{Answer Topic Analysis Word Tag Cloud}
    \label{fig:enter-label}
\end{figure*}

\begin{figure*}
    \centering
    \includegraphics[width=\linewidth]{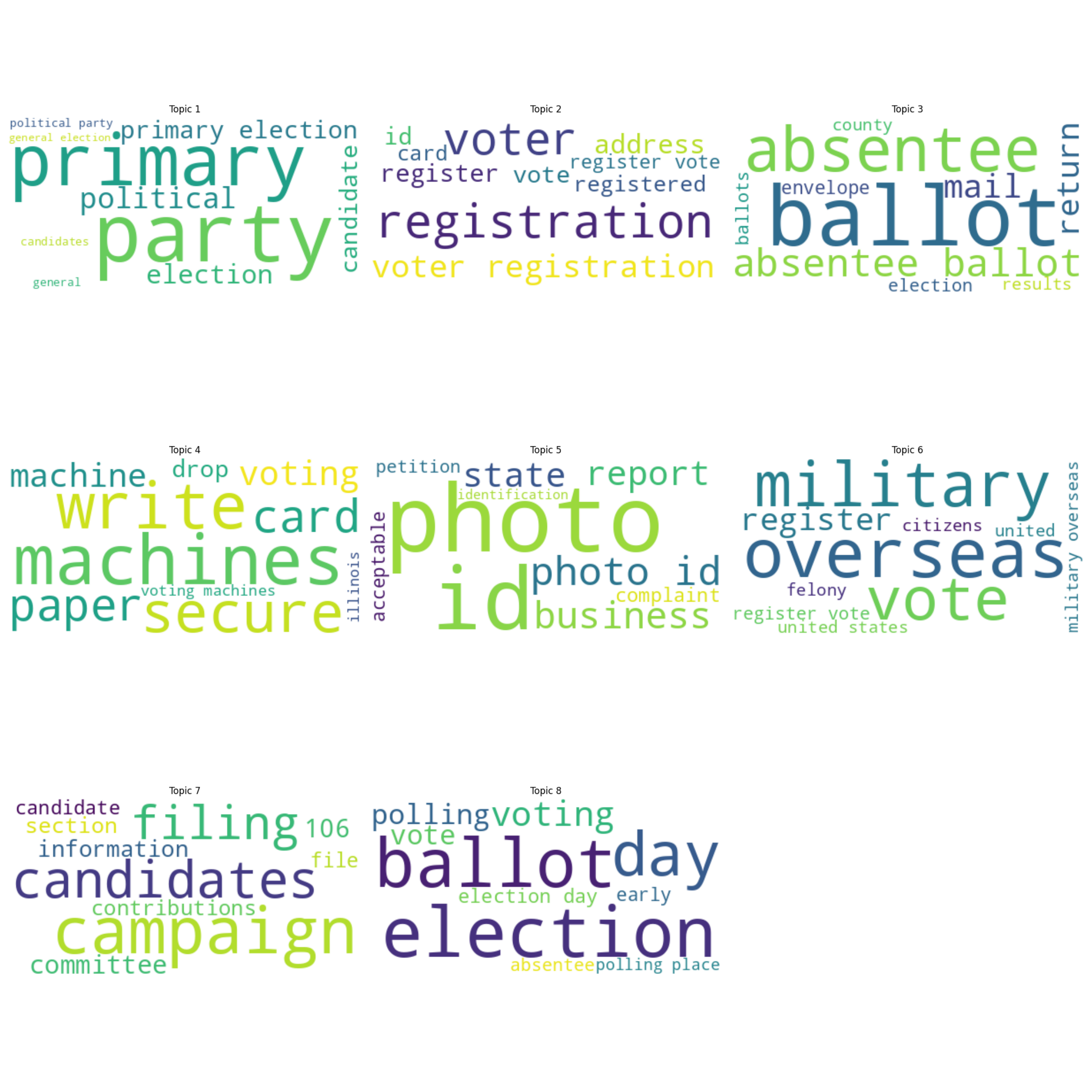}
    \caption{Question + Answer Topic Analysis Word Tag Cloud}
    \label{fig:enter-label}
\end{figure*}

\begin{table*}\centering
\caption{Question Topic Analysis}\label{tab:question_topic_analysis}
\scriptsize
\begin{tabular}{lrrrrrrrrrr}\toprule
\textbf{STATE} &\textbf{Topic 1} &\textbf{Topic 2} &\textbf{Topic 3} &\textbf{Topic 4} &\textbf{Topic 5} &\textbf{Topic 6} &\textbf{Topic 7} &\textbf{Topic 8} &\textbf{MEDIAN} \\\midrule
AK &0.14 &0.11 &0.11 &0.16 &0.10 &0.14 &0.12 &0.12 &0.12 \\
AL &0.23 &0.11 &0.10 &0.07 &0.13 &0.12 &0.12 &0.12 &0.12 \\
AR &0.16 &0.04 &0.27 &0.11 &0.09 &0.13 &0.14 &0.04 &0.12 \\
AZ &0.16 &0.09 &0.10 &0.14 &0.17 &0.08 &0.12 &0.15 &0.13 \\
CA &0.25 &0.12 &0.12 &0.07 &0.06 &0.07 &0.12 &0.19 &0.12 \\
CO &0.10 &0.10 &0.09 &0.11 &0.13 &0.08 &0.14 &0.25 &0.11 \\
CT &0.10 &0.24 &0.10 &0.08 &0.17 &0.09 &0.12 &0.12 &0.11 \\
\bottomrule
\end{tabular}
\end{table*}


\begin{table*}
\centering
\caption{Question Topic Analysis Summary}
\label{tab:q_topic_analysis_sum}
\scriptsize
\resizebox{\textwidth}{!}{%
\begin{tabular}{lrrrrrrr}\toprule
\textbf{Topic} &\textbf{Final Score} &\textbf{Prevalence} &\textbf{Coherence} &\textbf{Distinctiveness} &\textbf{Coverage} &\textbf{Top Terms} \\\midrule
Topic 1: Political Parties and Primary Elections &1.00 &1.00 &1.00 &1.00 &1.00 & registration, voter registration, voter, party, polling \\
Topic 2: Voter Registration &0.54 &0.87 &0.49 &0.14 &0.65 &vote, registered, register, registered vote, military \\
Topic 3: Absentee Voting &0.10 &0.00 &0.00 &0.39 &0.00 & register, register vote, vote, provisional, online \\
Topic 4: Voting Equipment Security &0.12 &0.98 &0.03 &0.26 &0.10 &vote, person, need, id, vote person \\
Topic 5: Voter Identification Requirements &0.25 &0.11 &0.06 &0.71 &0.11 &day, election day, ballots, election, polls \\
Topic 6: Military and Overseas Voting &0.11 &0.15 &0.14 &0.00 &0.14 &voting, early voting, ballot, absentee, early \\
Topic 7: Campaign Filing and Candidates &0.75 &0.92 &0.96 &0.20 &0.93 &ballot, mail, absentee, absentee ballot, mail ballot \\
Topic 8: Election Day and Polling Information &0.43 &0.40 &0.14 &0.71 &0.46 &election, primary, elections, primary election, general \\
\bottomrule
\end{tabular}%
}
\end{table*}

\begin{table*}\centering
\caption{Answer Topic Analysis}\label{tab:answer_topic_analysis}
\scriptsize
\begin{tabular}{lrrrrrrrrrr}\toprule
\textbf{STATE} &\textbf{Topic 1} &\textbf{Topic 2} &\textbf{Topic 3} &\textbf{Topic 4} &\textbf{Topic 5} &\textbf{Topic 6} &\textbf{Topic 7} &\textbf{Topic 8} &\textbf{MEDIAN} \\\midrule
AK &0.24 &0.07 &0.36 &0.03 &0.04 &0.07 &0.08 &0.13 &0.07 \\
AL &0.18 &0.15 &0.33 &0.03 &0.03 &0.09 &0.08 &0.10 &0.10 \\
AR &0.33 &0.03 &0.33 &0.03 &0.03 &0.06 &0.04 &0.17 &0.05 \\
AZ &0.30 &0.06 &0.32 &0.02 &0.03 &0.09 &0.10 &0.09 &0.09 \\
CA &0.35 &0.16 &0.20 &0.03 &0.03 &0.06 &0.06 &0.12 &0.09 \\
CO &0.33 &0.14 &0.28 &0.02 &0.03 &0.03 &0.10 &0.06 &0.08 \\
CT &0.17 &0.06 &0.35 &0.02 &0.04 &0.05 &0.20 &0.09 &0.08 \\
\bottomrule
\end{tabular}
\end{table*}


\begin{table*}\centering
\caption{Answer Topic Analysis Summary}\label{tab: a_topic_analysis_sum}
\scriptsize
\resizebox{\textwidth}{!}{%
\begin{tabular}{lrrrrrrr}\toprule
\textbf{Topic} &\textbf{Final Score} &\textbf{Prevalence} &\textbf{Coherence} &\textbf{Distinctiveness} &\textbf{Coverage} &\textbf{Top Terms} \\\midrule
Topic 1: Political Parties and Primary Elections &0.65 &0.77 &0.78 &0.18 &0.85 & registration, voter, voter registration, address, county \\
Topic 2: Voter Registration &0.46 &0.22 &0.36 &1.00 &0.28 &party, primary, candidates, political, candidate \\
Topic 3: Absentee Voting &0.75 &1.00 &1.00 &0.00 &1.00 &ballot, absentee, voting, election, vote \\
Topic 4: Voting Equipment Security &0.01 &0.00 &0.00 &0.21 &0.00 & Louisiana, majority, la 18, view, la \\
Topic 5: Voter Identification Requirements &0.09 &0.05 &0.06 &0.16 &0.07 &military, overseas, united states, united, military overseas \\
Topic 6: Military and Overseas Voting &0.18 &0.12 &0.10 &0.35 &0.14 &campaign, overseas, filing, contributions \\
Topic 7: Campaign Filing and Candidates &0.40 &0.20 &0.27 &0.84 &0.30 &id, identification, photo, license, photo id \\
Topic 8: Election Day and Polling Information &0.26 &0.29 &0.19 &0.15 &0.40 &state, secretary state, election, elections, secretary \\
\bottomrule
\end{tabular}%
}
\end{table*}

\begin{table*}\centering
\caption{Question + Answer Topic Analysis}\label{tab:qa_topic_analysis}
\scriptsize
\begin{tabular}{lrrrrrrrrrr}\toprule
\textbf{STATE} &\textbf{Topic 1} &\textbf{Topic 2} &\textbf{Topic 3} &\textbf{Topic 4} &\textbf{Topic 5} &\textbf{Topic 6} &\textbf{Topic 7} &\textbf{Topic 8} &\textbf{MEDIAN} \\\midrule
AK &0.08 &0.20 &0.20 &0.04 &0.06 &0.09 &0.06 &0.26 &0.09 \\
AL &0.19 &0.16 &0.12 &0.04 &0.04 &0.07 &0.10 &0.28 &0.11 \\
AR &0.02 &0.29 &0.06 &0.06 &0.03 &0.15 &0.05 &0.33 &0.06 \\
AZ &0.11 &0.19 &0.10 &0.04 &0.05 &0.07 &0.07 &0.37 &0.09 \\
CA &0.16 &0.34 &0.08 &0.02 &0.09 &0.04 &0.04 &0.22 &0.09 \\
CO &0.16 &0.17 &0.07 &0.05 &0.10 &0.02 &0.03 &0.41 &0.08 \\
CT &0.06 &0.23 &0.09 &0.03 &0.08 &0.15 &0.03 &0.33 &0.09 \\
\bottomrule
\end{tabular}
\end{table*}

\begin{table*}\centering
\caption{Question + Answer Topic Analysis Summary}\label{tab:qa_topic_analysis_sum }
\scriptsize
\resizebox{\textwidth}{!}{%
\begin{tabular}{lrrrrrrr}\toprule
\textbf{Topic} &\textbf{Final Score} &\textbf{Prevalence} &\textbf{Coherence} &\textbf{Distinctiveness} &\textbf{Coverage} &\textbf{Top Terms} \\\midrule
Topic 1: Political Parties and Primary Elections &0.46 &0.28 &0.44 &0.80 &308 &party, primary, political, election, primary election \\
Topic 2: Voter Registration &0.76 &0.82 &0.91 &0.52 &0.79 & registration, voter, voter registration, address, register \\
Topic 3: Absentee Voting &0.41 &0.49 &0.53 &0.08 &0.52 &ballot, absentee, absentee ballot, mail, return \\
Topic 4: Voting Equipment Security &0.16 &0.00 &0.00 &0.65 &0.00 &machines, write, secure, paper, card \\
Topic 5: Voter Identification Requirements &0.24 &0.06 &0.06 &0.69 &0.13 &photo, id, photo id, business, report \\
Topic 6: Military and Overseas Voting &0.39 &0.16 &0.19 &1.00 &0.22 &overseas, military, vote, register, register vote \\
Topic 7: Campaign Filing and Candidates &0.26 &0.06 &0.08 &0.77 &0.11 &campaign, candidates, filing, committee \\
Topic 8: Election Day and Polling Information &0.75 &1.00 &1.00 &0.00 &1.00 &election, ballot, day, voting, polling \\
\bottomrule
\end{tabular}%
}
\end{table*}

 
\begin{table*}\centering
\caption{Question Sentiment Analysis}\label{tab:q_sentiment_analysis}
\scriptsize
\resizebox{\textwidth}{!}{%
\begin{tabular}{lrrrrrrrrrrr}\toprule
State &Average Compound Score &Positive Count &Negative Count &Neutral Count &Question Positive \% &Question Negative \% &Question Neutral \% &Positive Compound Score &Negative Compound Score &Neutral Compound Score \\\midrule
AK &-0.09 &1.00 &23.00 &31.00 &1.82 &41.82 &56.36 &0.79 &-5.82 &0.00 \\
AL &0.02 &6.00 &3.00 &33.00 &14.29 &7.14 &78.57 &1.63 &-0.89 &0.00 \\
AR &-0.05 &0.00 &4.00 &18.00 &0.00 &18.18 &81.82 &0.00 &-1.09 &0.00 \\
AZ &-0.01 &4.00 &4.00 &27.00 &11.43 &11.43 &77.14 &1.17 &-1.46 &0.00 \\
CA &0.01 &10.00 &6.00 &64.00 &12.50 &7.50 &80.00 &2.96 &-1.77 &0.00 \\
CO &0.01 &4.00 &2.00 &39.00 &8.89 &4.44 &86.67 &1.18 &-0.69 &0.00 \\
CT &-0.03 &3.00 &9.00 &47.00 &5.08 &15.25 &79.66 &0.95 &-2.86 &0.00 \\
\bottomrule
\end{tabular}%
}
\end{table*}

\begin{table*}\centering
\caption{Answer Sentiment Analysis}\label{tab:a_sentiment_analysis}
\scriptsize
\resizebox{\textwidth}{!}{%
\begin{tabular}{lrrrrrrrrrrr}\toprule
State &Average Compund Score &Positive Count &Negative Count &Neutral Count &Answer Positive \% &Answer Negative \% &Answer Neutral \% &Positive Compound Score &Negative Compound Score &Neutral Compound Score \\\midrule
AK &0.00 &22.00 &24.00 &9.00 &40.00 &43.64 &16.36 &10.34 &-10.58 &0.00 \\
AL &0.21 &25.00 &6.00 &11.00 &59.52 &14.29 &26.19 &12.17 &-3.39 &0.00 \\
AR &0.03 &8.00 &7.00 &7.00 &36.36 &31.82 &31.82 &4.45 &-3.78 &0.00 \\
AZ &0.34 &22.00 &6.00 &7.00 &62.86 &17.14 &20.00 &13.38 &-1.64 &0.00 \\
CA &0.27 &53.00 &14.00 &13.00 &66.25 &17.50 &16.25 &28.44 &-6.61 &0.00 \\
CO &0.15 &22.00 &13.00 &10.00 &48.89 &28.89 &22.22 &12.74 &-5.78 &0.00 \\
CT &0.07 &25.00 &21.00 &13.00 &42.37 &35.59 &22.03 &13.31 &-9.32 &0.00 \\
\bottomrule
\end{tabular}%
}
\end{table*}

\begin{table*}\centering
\caption{Question + Answer Sentiment Analysis}\label{tab:qa_sentiment_analysis}
\scriptsize
\resizebox{\textwidth}{!}{%
\begin{tabular}{lrrrrrrrrrrr}\toprule
State &Average Sentiment Score &Positive Count &Negative Count &Neutral Count &Positive \% &Negative \% &Neutral \% &Positive Compound Score &Negative Compound Score &Neutral Compound Score \\\midrule
AK &-0.07 &21.00 &29.00 &5.00 &38.18 &52.73 &9.09 &9.82 &-13.61 &0.00 \\
AL &0.22 &27.00 &7.00 &8.00 &64.29 &16.67 &19.05 &13.18 &-3.80 &0.00 \\
AR &0.01 &8.00 &7.00 &7.00 &36.36 &31.82 &31.82 &4.45 &-4.19 &0.00 \\
AZ &0.33 &24.00 &6.00 &5.00 &68.57 &17.14 &14.29 &13.86 &-2.40 &0.00 \\
CA &0.27 &51.00 &16.00 &13.00 &63.75 &20.00 &16.25 &28.91 &-7.10 &0.00 \\
CO &0.16 &22.00 &13.00 &10.00 &48.89 &28.89 &22.22 &13.24 &-5.88 &0.00 \\
CT &0.05 &23.00 &24.00 &12.00 &38.98 &40.68 &20.34 &13.72 &-10.79 &0.00 \\
\bottomrule
\end{tabular}%
}
\end{table*}

\begin{figure*}
     \centering
     \includegraphics[width=\textwidth]{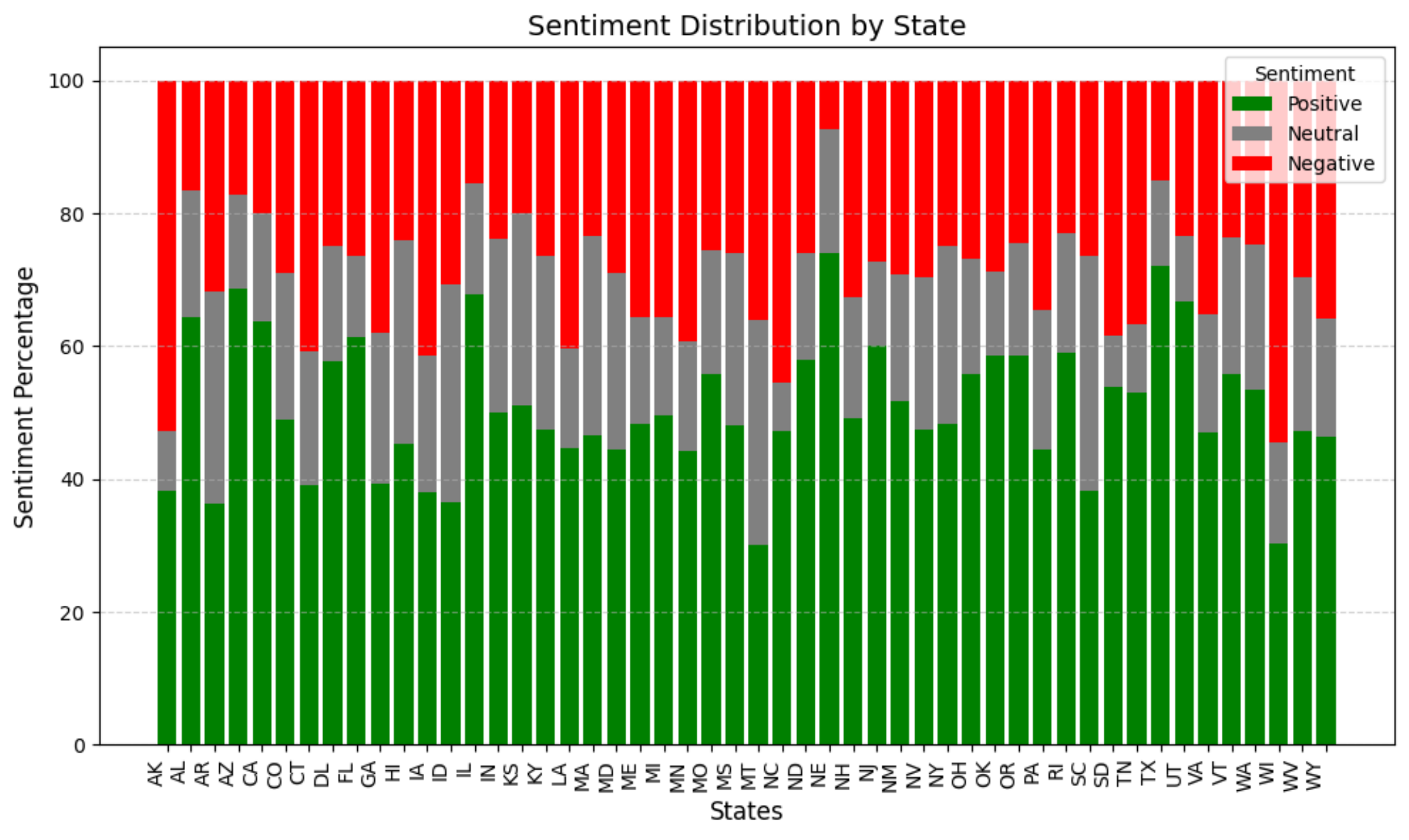}
     \caption{Distribution of sentiments.}
     \label{fig:senti-distrib}
 \end{figure*}

 \begin{figure*}
     \centering
     \includegraphics[width=\textwidth]{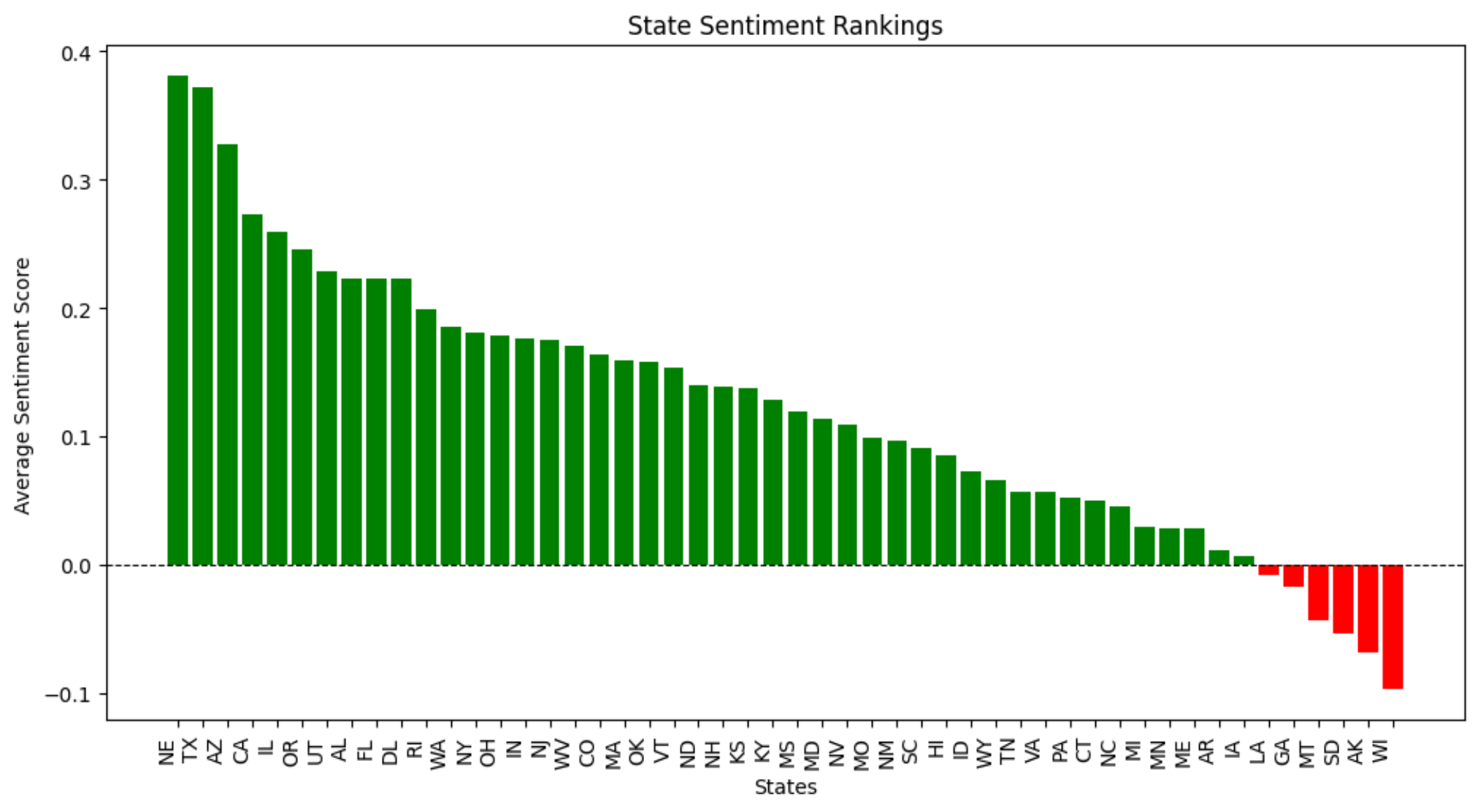}
     \caption{Range of sentiments.}
     \label{fig:senti-range}
 \end{figure*}

\begin{table*}\centering
\caption{LLM Answer Readability Analysis}\label{tab:llm_readability_analysis_a}
\scriptsize
\begin{tabular}{lrrrrrrr}\toprule
State &FKG &GFI &SI &ARI &CLI &AVERAGE \\\midrule
AK &10.36 &11.17 &7.40 &12.10 &10.85 &10.37 \\
AL &9.93 &10.73 &8.06 &12.08 &11.51 &10.46 \\
AR &12.37 &14.40 &7.40 &14.52 &10.04 &11.74 \\
AZ &9.51 &10.42 &9.35 &11.49 &11.41 &10.44 \\
CA &11.49 &11.94 &9.17 &14.32 &12.53 &11.89 \\
CO &9.68 &10.03 &7.51 &12.90 &12.64 &10.55 \\
CT &10.29 &11.39 &9.03 &12.29 &10.94 &10.79 \\
\bottomrule
\end{tabular}
\end{table*}

\begin{table*}\centering
\caption{LLM Question + Answer readability Analysis}\label{tab:llm_readability_analysis_qa}
\scriptsize
\begin{tabular}{lrrrrrrr}\toprule
STATE &FKG &GFI &SI &ARI &CLI &Average \\\midrule
AK &9.30 &9.83 &9.39 &10.43 &10.42 &9.88 \\
AL &8.98 &9.40 &9.38 &10.34 &10.61 &9.74 \\
AR &9.85 &11.34 &8.45 &11.14 &9.57 &10.07 \\
AZ &9.13 &9.75 &9.85 &10.75 &11.08 &10.11 \\
CA &10.72 &10.78 &10.45 &13.00 &12.46 &11.48 \\
CO &8.90 &9.03 &9.79 &11.46 &11.72 &10.18 \\
CT &9.06 &9.94 &10.23 &10.47 &10.08 &9.95 \\
\bottomrule
\end{tabular}
\end{table*}

\begin{figure*}
    \centering
    \includegraphics[width=\textwidth]{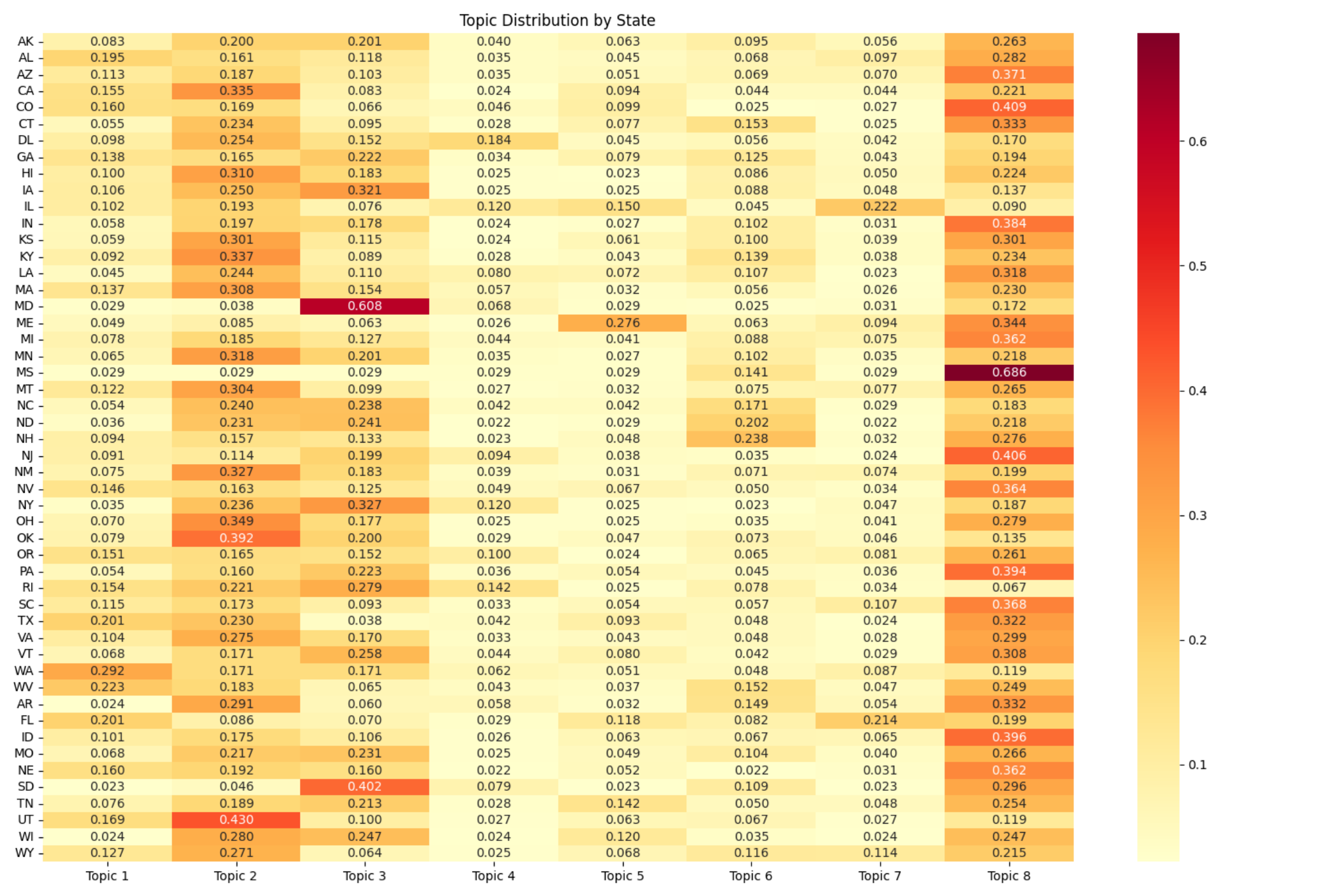} 
    \caption{This heatmap visualizes the distribution of eight election-related topics across U.S. states. Darker colors indicate stronger topic representation, while lighter colors highlight weaker coverage. The gradient underscores variations in FAQ coverage across states, with rows representing states (postal abbreviations) and columns the identified topics.}
    \label{fig:topic_distribution_heatmap}
\end{figure*}

\begin{figure*}
    \centering
    \includegraphics[width=\linewidth]{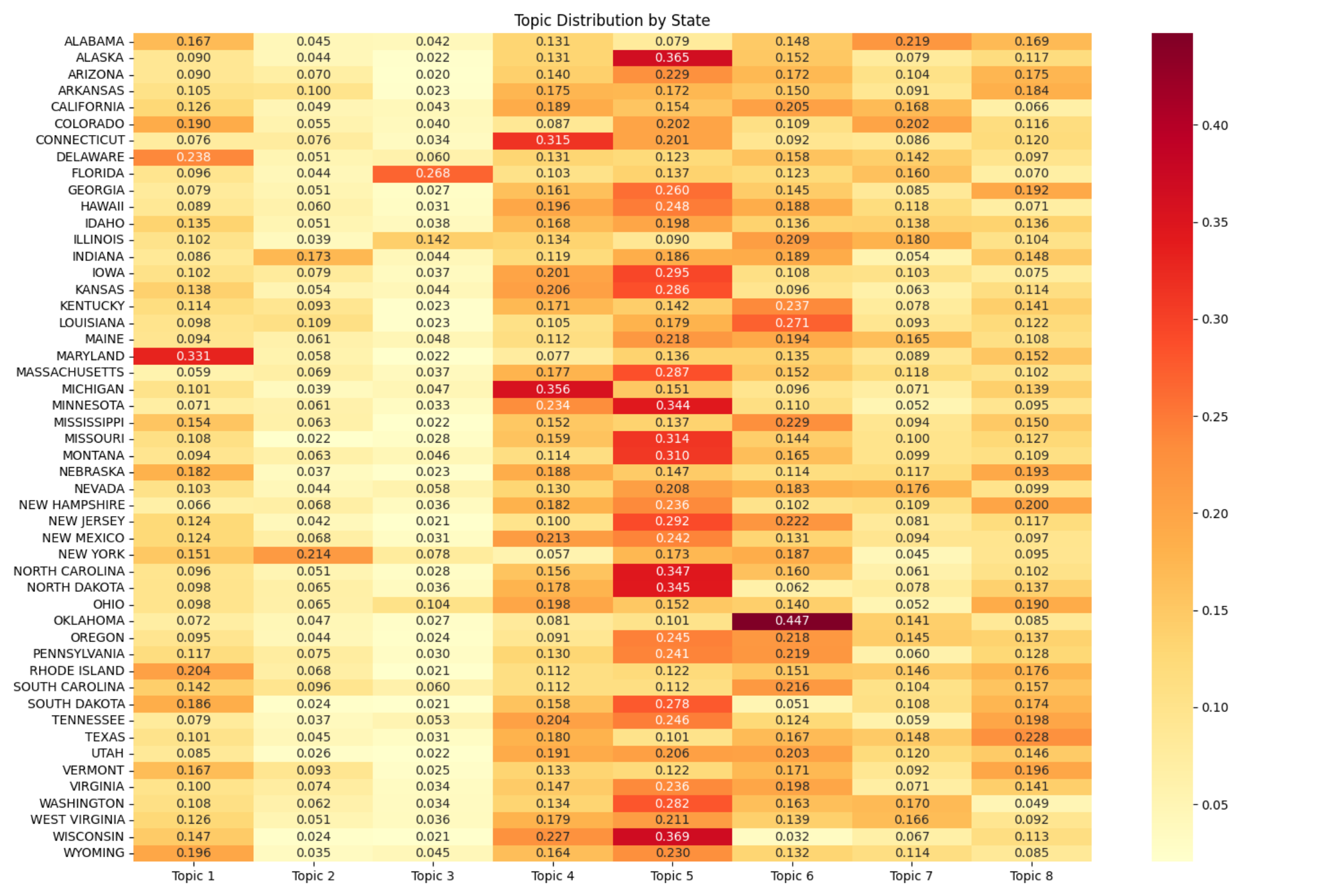}
    \caption{This heatmap visualizes the distribution of eight election-related topics across U.S. states. Darker colors indicate stronger topic representation, while lighter colors highlight weaker coverage. The gradient underscores variations in FAQ coverage across states, with rows representing states (postal abbreviations) and columns the identified topics on the generated answers}
    \label{fig:state_topic_dist}
\end{figure*}

\begin{table*}\centering
\caption{LLM Abstractive Question Summary Analysis}\label{tab:llm_a_summary_analysis_q}
\scriptsize
\begin{tabular}{lrrrrrrrrrr}\toprule
State &Cosine\_Similarity &ROUGE\_1 &ROUGE\_2 &ROUGE\_L &ROUGE\_W &ROUGE\_S &ROUGE\_SU &BLEU\_Score &Semantic Overlap \\\midrule
AK &0.55 &0.45 &0.35 &0.39 &0.34 &0.29 &0.32 &0.16 &0.50 \\
AL &0.67 &0.54 &0.44 &0.48 &0.41 &0.36 &0.39 &0.22 &0.61 \\
AR &0.61 &0.51 &0.41 &0.45 &0.37 &0.31 &0.35 &0.18 &0.56 \\
AZ &0.59 &0.47 &0.38 &0.41 &0.35 &0.31 &0.34 &0.16 &0.53 \\
CA &0.60 &0.50 &0.39 &0.42 &0.35 &0.31 &0.34 &0.17 &0.55 \\
CO &0.60 &0.46 &0.35 &0.38 &0.32 &0.28 &0.31 &0.15 &0.53 \\
CT &0.56 &0.48 &0.37 &0.41 &0.36 &0.30 &0.33 &0.19 &0.52 \\
\bottomrule
\end{tabular}
\end{table*}

\begin{table*}\centering
\caption{LLM Extractive Question Summary Analysis}\label{tab:llm_e_summary_analysis_q}
\scriptsize
\begin{tabular}{lrrrrrrrrrr}\toprule
State &Cosine\_Similarity &ROUGE\_1 &ROUGE\_2 &ROUGE\_L &ROUGE\_W &ROUGE\_S &ROUGE\_SU &BLEU\_Score &Semantic Overlap \\\midrule
AK &0.03 &0.05 &0.01 &0.05 &0.02 &0.00 &0.01 &0.00 &0.04 \\
AL &0.07 &0.07 &0.02 &0.06 &0.05 &0.01 &0.02 &0.01 &0.07 \\
AR &0.04 &0.05 &0.01 &0.04 &0.01 &0.00 &0.00 &0.00 &0.04 \\
AZ &0.06 &0.09 &0.03 &0.08 &0.06 &0.01 &0.02 &0.01 &0.07 \\
CA &0.08 &0.11 &0.03 &0.09 &0.06 &0.01 &0.03 &0.01 &0.10 \\
CO &0.07 &0.09 &0.04 &0.07 &0.06 &0.02 &0.03 &0.02 &0.08 \\
CT &0.04 &0.05 &0.01 &0.05 &0.03 &0.00 &0.01 &0.00 &0.04 \\
\bottomrule
\end{tabular}
\end{table*}

\begin{table*}\centering
\caption{LLM Abstractive Answer Summary Analysis}\label{tab:llm_a_summary_analysis_a}
\scriptsize
\begin{tabular}{lrrrrrrrrrr}\toprule
State &Cosine\_Similarity &ROUGE\_1 &ROUGE\_2 &ROUGE\_L &ROUGE\_W &ROUGE\_S &ROUGE\_SU &BLEU\_Score &Semantic Overlap \\\midrule
AK &0.55 &0.45 &0.35 &0.39 &0.34 &0.29 &0.32 &0.16 &0.50 \\
AL &0.67 &0.54 &0.44 &0.48 &0.41 &0.36 &0.39 &0.22 &0.61 \\
AR &0.61 &0.51 &0.41 &0.45 &0.37 &0.31 &0.35 &0.18 &0.56 \\
AZ &0.59 &0.47 &0.38 &0.41 &0.35 &0.31 &0.34 &0.16 &0.53 \\
CA &0.60 &0.50 &0.39 &0.42 &0.35 &0.31 &0.34 &0.17 &0.55 \\
CO &0.60 &0.46 &0.35 &0.38 &0.32 &0.28 &0.31 &0.15 &0.53 \\
CT &0.56 &0.48 &0.37 &0.41 &0.36 &0.30 &0.33 &0.19 &0.52 \\
\bottomrule
\end{tabular}
\end{table*}

\begin{table*}\centering
\caption{LLM Extractive Answer Summary Analysis}\label{tab:llm_e_summary_analysis_a}
\scriptsize
\begin{tabular}{lrrrrrrrrrr}\toprule
State &Cosine\_Similarity &ROUGE\_1 &ROUGE\_2 &ROUGE\_L &ROUGE\_W &ROUGE\_S &ROUGE\_SU &BLEU\_Score &Semantic Overlap \\\midrule
AK &0.16 &0.08 &0.06 &0.07 &0.06 &0.06 &0.06 &0.00 &0.12 \\
AL &0.22 &0.14 &0.11 &0.12 &0.11 &0.11 &0.11 &0.03 &0.18 \\
AR &0.13 &0.09 &0.06 &0.08 &0.07 &0.07 &0.07 &0.04 &0.11 \\
AZ &0.15 &0.07 &0.06 &0.07 &0.05 &0.05 &0.06 &0.00 &0.11 \\
CA &0.28 &0.17 &0.13 &0.15 &0.13 &0.12 &0.13 &0.02 &0.22 \\
CO &0.26 &0.15 &0.11 &0.13 &0.11 &0.10 &0.11 &0.02 &0.20 \\
CT &0.09 &0.04 &0.02 &0.03 &0.03 &0.02 &0.02 &0.00 &0.07 \\
\bottomrule
\end{tabular}
\end{table*}

\begin{table*}\centering
\caption{LLM Abstractive Question + Answer Summary Analysis}\label{tab:llm_a_summary_analysis_qa}
\scriptsize
\begin{tabular}{lrrrrrrrrrr}\toprule
State &Cosine\_Similarity &ROUGE\_1 &ROUGE\_2 &ROUGE\_L &ROUGE\_W &ROUGE\_S &ROUGE\_SU &BLEU\_Score &Semantic Overlap \\\midrule
AK &0.75 &0.67 &0.60 &0.60 &0.53 &0.51 &0.54 &0.38 &0.71 \\
AL &0.84 &0.72 &0.69 &0.70 &0.61 &0.57 &0.60 &0.43 &0.78 \\
AR &0.77 &0.66 &0.61 &0.60 &0.52 &0.50 &0.52 &0.36 &0.71 \\
AZ &0.72 &0.62 &0.55 &0.55 &0.48 &0.47 &0.50 &0.31 &0.67 \\
CA &0.78 &0.66 &0.61 &0.63 &0.55 &0.52 &0.55 &0.34 &0.72 \\
CO &0.76 &0.63 &0.56 &0.57 &0.50 &0.47 &0.49 &0.33 &0.70 \\
CT &0.74 &0.65 &0.59 &0.59 &0.52 &0.50 &0.53 &0.37 &0.70 \\
\bottomrule
\end{tabular}
\end{table*}

\begin{table*}\centering
\caption{LLM Extractive Question + Answer Summary Analysis}\label{tab:llm_e_summary_analysis_qa}
\scriptsize
\begin{tabular}{lrrrrrrrrrr}\toprule
State &Cosine\_Similarity &ROUGE\_1 &ROUGE\_2 &ROUGE\_L &ROUGE\_W &ROUGE\_S &ROUGE\_SU &BLEU\_Score &Semantic Overlap \\\midrule
AK &0.31 &0.22 &0.19 &0.19 &0.16 &0.18 &0.20 &0.07 &0.27 \\
AL &0.31 &0.22 &0.20 &0.21 &0.19 &0.20 &0.21 &0.08 &0.26 \\
AR &0.36 &0.30 &0.26 &0.26 &0.24 &0.26 &0.27 &0.18 &0.33 \\
AZ &0.37 &0.25 &0.21 &0.22 &0.19 &0.19 &0.21 &0.05 &0.31 \\
CA &0.45 &0.31 &0.28 &0.28 &0.25 &0.28 &0.30 &0.09 &0.38 \\
CO &0.42 &0.30 &0.26 &0.25 &0.23 &0.25 &0.27 &0.10 &0.36 \\
CT &0.38 &0.28 &0.24 &0.24 &0.22 &0.23 &0.24 &0.10 &0.33 \\
\bottomrule
\end{tabular}
\end{table*}

\begin{table*}\centering
\caption{LLM Answer Topic Analysis}\label{tab:llm_topic_analysis_a}
\scriptsize
\begin{tabular}{lrrrrrrrrrrr}\toprule
\textbf{STATE} &\textbf{Topic 1} &\textbf{Topic 2} &\textbf{Topic 3} &\textbf{Topic 4} &\textbf{Topic 5} &\textbf{Topic 6} &\textbf{Topic 7} &\textbf{Topic 8} &\textbf{Median} &\textbf{Average} \\\midrule
AK &0.08 &0.24 &0.10 &0.05 &0.07 &0.08 &0.28 &0.09 &0.09 &0.13 \\
AL &0.09 &0.17 &0.22 &0.07 &0.08 &0.09 &0.13 &0.14 &0.11 &0.13 \\
AR &0.06 &0.29 &0.11 &0.06 &0.11 &0.14 &0.14 &0.10 &0.11 &0.13 \\
AZ &0.05 &0.29 &0.12 &0.07 &0.09 &0.09 &0.23 &0.06 &0.09 &0.13 \\
CA &0.05 &0.31 &0.16 &0.06 &0.11 &0.09 &0.13 &0.09 &0.10 &0.13 \\
CO &0.06 &0.19 &0.17 &0.06 &0.11 &0.08 &0.25 &0.09 &0.10 &0.13 \\
CT &0.12 &0.30 &0.10 &0.05 &0.09 &0.11 &0.13 &0.09 &0.11 &0.13 \\
\bottomrule
\end{tabular}
\end{table*}

\begin{table*}\centering
\caption{LLM Answer Topic Analysis Summary}\label{tab:llm_topic_analysis_sum_a}
\scriptsize
\begin{tabular}{lrrrrrrr}\toprule
\textbf{Topic} &\textbf{Final Score} &\textbf{Prevalence} &\textbf{Coherence} &\textbf{Distinctiveness} &\textbf{Coverage} &\textbf{Top Terms} \\\midrule
Topic 1 &0.20 &0.16 &0.08 &0.36 &0.22 & registration, michigan, vote, time, register \\
Topic 2 &0.75 &1.00 &0.96 &0.05 &1.00 &voter, registration, id, voter registration, vote \\
Topic 3 &0.49 &0.33 &0.42 &0.87 &0.36 &primary, party, states, questions, 2024 \\
Topic 4 &0.25 &0.00 &0.00 &1.00 &0.00 &results, election, available, secretary, measures \\
Topic 5 &0.34 &0.18 &0.16 &0.74 &0.27 &00, overseas, 00 00, military, open \\
Topic 6 &0.23 &0.24 &0.02 &0.37 &0.30 &vote411, www vote411, www, vote411 org, elections \\
Topic 7 &0.66 &0.80 &1.00 &0.00 &0.85 &ballot, absentee, election, county, mail \\
Topic 8 &0.24 &0.31 &0.12 &0.16 &0.37 &poll, ballot, voting, worker, election \\
\bottomrule
\end{tabular}
\end{table*}

\begin{table*}\centering
\caption{LLM Question + Answer Topic Analysis}\label{tab:llm_topic_analysis_qa}
\scriptsize
\begin{tabular}{lrrrrrrrrrrr}\toprule
\textbf{STATE} &\textbf{Topic 1} &\textbf{Topic 2} &\textbf{Topic 3} &\textbf{Topic 4} &\textbf{Topic 5} &\textbf{Topic 6} &\textbf{Topic 7} &\textbf{Topic 8} &\textbf{Median} &\textbf{Average} \\\midrule
AK &0.44 &0.05 &0.16 &0.12 &0.04 &0.10 &0.05 &0.04 &0.08 &0.13 \\
AL &0.20 &0.09 &0.30 &0.11 &0.04 &0.12 &0.08 &0.05 &0.10 &0.13 \\
AR &0.33 &0.08 &0.14 &0.08 &0.07 &0.17 &0.07 &0.06 &0.08 &0.13 \\
AZ &0.42 &0.08 &0.15 &0.12 &0.04 &0.09 &0.06 &0.04 &0.09 &0.13 \\
CA &0.38 &0.07 &0.27 &0.07 &0.04 &0.07 &0.07 &0.03 &0.07 &0.13 \\
CO &0.43 &0.04 &0.22 &0.08 &0.04 &0.07 &0.08 &0.04 &0.07 &0.13 \\
CT &0.37 &0.07 &0.14 &0.18 &0.04 &0.12 &0.05 &0.02 &0.10 &0.13 \\
\bottomrule
\end{tabular}
\end{table*}


\begin{figure*}
    \centering
    \includegraphics[width=\linewidth]{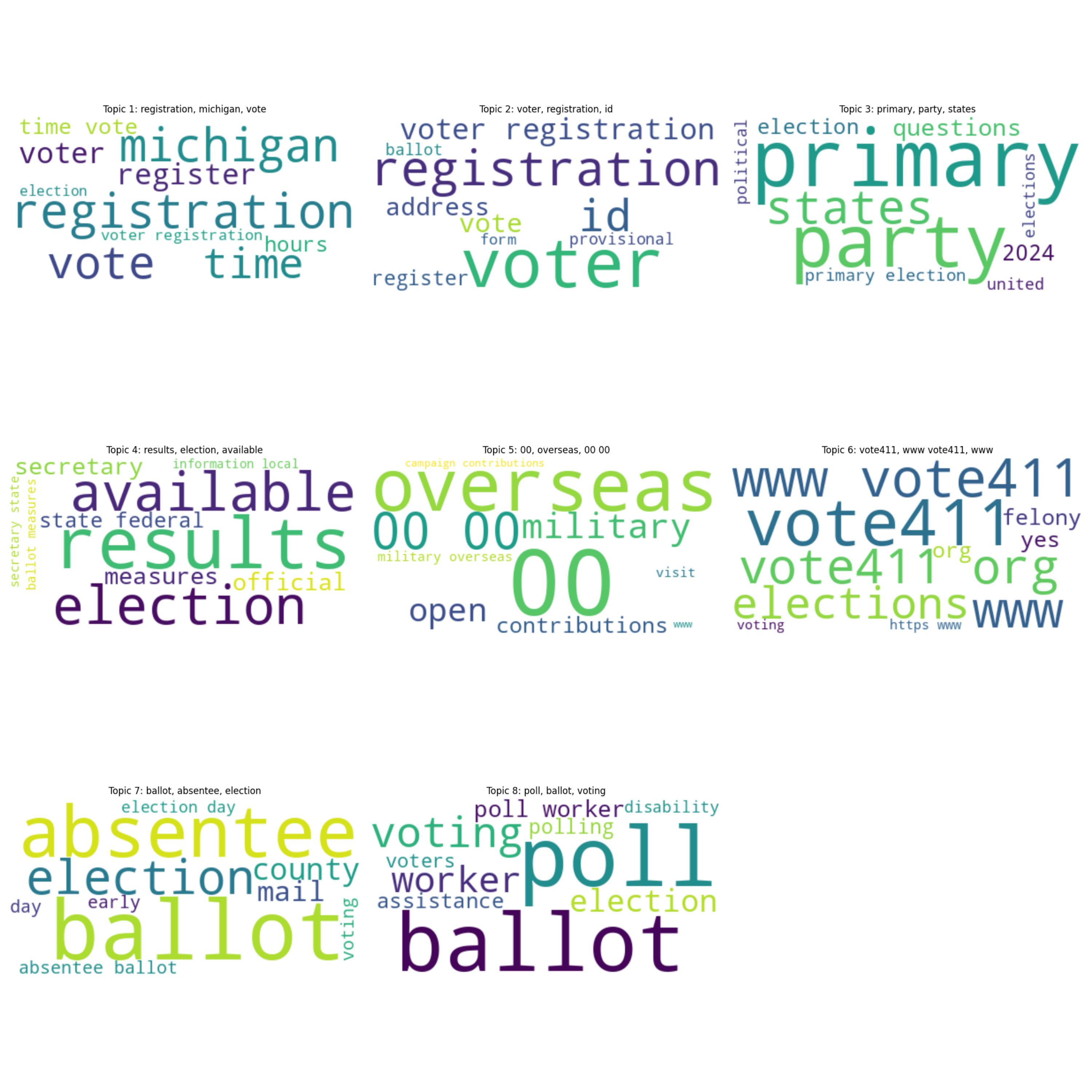}
    \caption{LLM Answer Topic Analysis Word Tag Cloud}
    \label{fig:enter-label}
\end{figure*}

\begin{figure*}
    \centering
    \includegraphics[width=\linewidth]{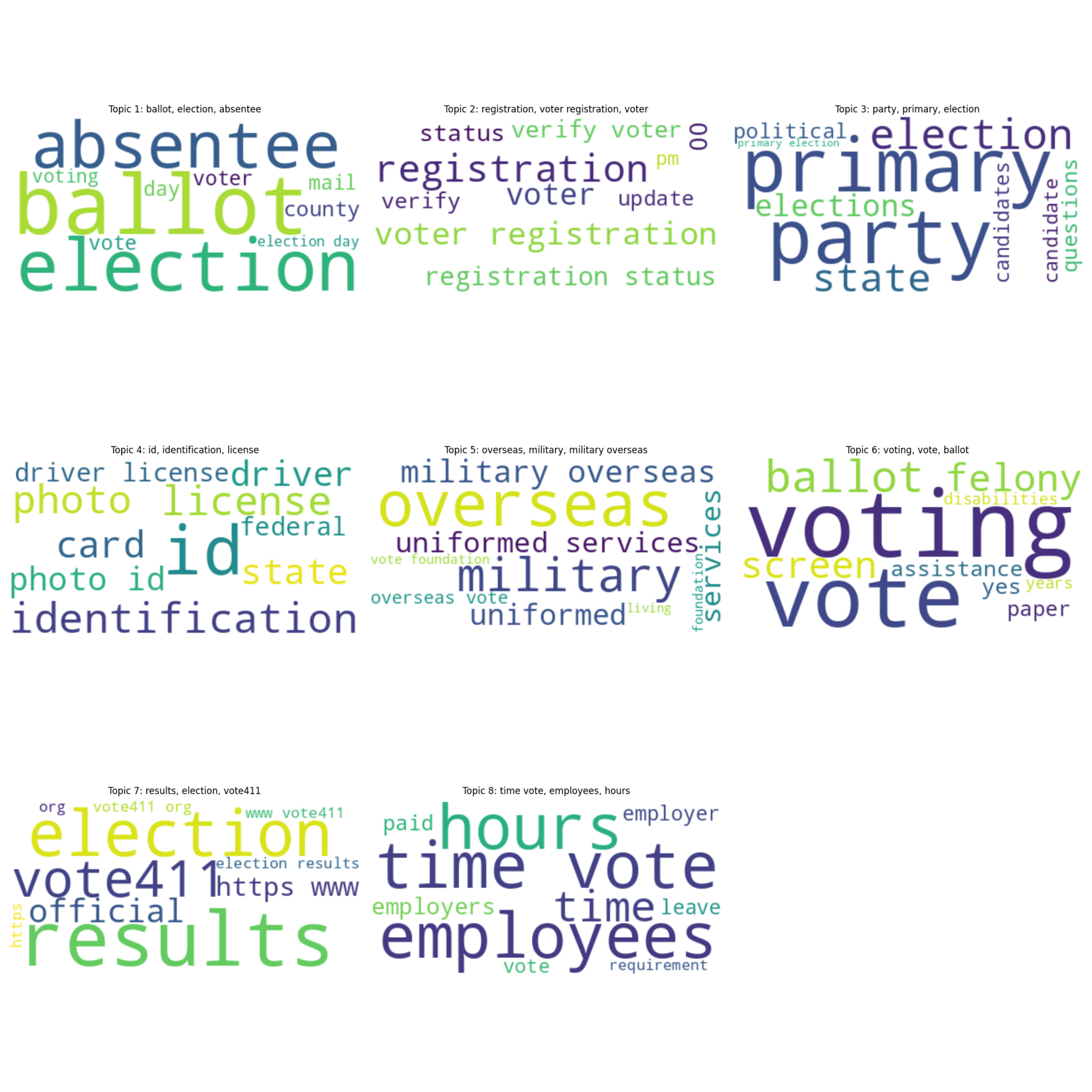}
    \caption{LLM Question + Answer Topic Analysis Word Tag Cloud}
    \label{fig:enter-label}
\end{figure*}

\begin{table*}\centering
\caption{LLM Question + Answer Topic Analysis Summary}\label{tab:llm_topic_analysis_sum_qa}
\scriptsize
\begin{tabular}{lrrrrrrr}\toprule
\textbf{Topic} &\textbf{Final Score} &\textbf{Prevalence} &\textbf{Coherence} &\textbf{Distinctiveness} &\textbf{Coverage} &\textbf{Top Terms} \\\midrule
Topic 1 &0.75 &1.00 &1.00 &0.00 &1.00 &ballot, election, absentee, county, voter \\
Topic 2 &0.13 &0.10 &0.16 &0.13 &0.15 &  registration, voter registration, voter, regis... \\
Topic 3 &0.39 &0.47 &0.34 &0.21 &0.52 &party, primary, election, state, elections \\
Topic 4 &0.33 &0.18 &0.20 &0.67 &0.25 &id, identification, license, photo, card \\
Topic 5 &0.27 &0.02 &0.04 &1.00 &0.03 &overseas, military, military overseas, uniform... \\
Topic 6 &0.19 &0.21 &0.11 &0.19 &0.25 &voting, vote, ballot, felony, screen \\
Topic 7 &0.21 &0.05 &0.03 &0.70 &0.06 &results, election, vote411, official \\
Topic 8 &0.25 &0.00 &0.00 &1.00 &0.00 &time vote, employees, hours, time, employers \\
\bottomrule
\end{tabular}
\end{table*}

\begin{table*}\centering
\caption{LLM Answer Sentiment Analysis}\label{tab:llm_sentiment_analysis_a}
\scriptsize
\begin{tabular}{lrrrrrrrrr}\toprule
State &Total Compound &Positive &Negative &Neutral &Positive \% &Negative \% &Neutral \% &Average Sentiment Score \\\midrule
AK &3.63 &23.00 &16.00 &16.00 &41.82 &29.09 &29.09 &0.07 \\
AL &10.84 &25.00 &6.00 &11.00 &59.52 &14.29 &26.19 &0.26 \\
AR &2.31 &12.00 &5.00 &5.00 &54.55 &22.73 &22.73 &0.11 \\
AZ &13.28 &24.00 &5.00 &6.00 &68.57 &14.29 &17.14 &0.38 \\
CA &27.22 &57.00 &10.00 &13.00 &71.25 &12.50 &16.25 &0.34 \\
CO &13.71 &30.00 &7.00 &8.00 &66.67 &15.56 &17.78 &0.31 \\
CT &7.45 &30.00 &16.00 &13.00 &50.85 &27.12 &22.03 &0.13 \\
DE &28.80 &59.00 &18.00 &27.00 &56.73 &17.31 &25.96 &0.28 \\
FL &41.74 &84.00 &24.00 &24.00 &63.64 &18.18 &18.18 &0.32 \\
GA &4.33 &30.00 &30.00 &19.00 &37.97 &37.97 &24.05 &0.06 \\
HI &14.25 &37.00 &14.00 &24.00 &49.33 &18.67 &32.00 &0.19 \\
IA &8.09 &30.00 &17.00 &11.00 &51.72 &29.31 &18.97 &0.14 \\
ID &4.05 &21.00 &14.00 &17.00 &40.38 &26.92 &32.69 &0.08 \\
IL &24.48 &60.00 &11.00 &19.00 &66.67 &12.22 &21.11 &0.27 \\
IN &12.76 &42.00 &13.00 &25.00 &52.50 &16.25 &31.25 &0.16 \\
KS &8.36 &27.00 &8.00 &10.00 &60.00 &17.78 &22.22 &0.19 \\
KY &4.61 &24.00 &16.00 &17.00 &42.11 &28.07 &29.82 &0.08 \\
LA &10.91 &56.00 &39.00 &19.00 &49.12 &34.21 &16.67 &0.10 \\
MA &15.34 &46.00 &19.00 &25.00 &51.11 &21.11 &27.78 &0.17 \\
MD &12.37 &25.00 &9.00 &11.00 &55.56 &20.00 &24.44 &0.28 \\
ME &5.83 &30.00 &18.00 &8.00 &53.57 &32.14 &14.29 &0.10 \\
MI &20.26 &87.00 &46.00 &24.00 &55.41 &29.30 &15.29 &0.13 \\
MN &4.80 &34.00 &28.00 &17.00 &43.04 &35.44 &21.52 &0.06 \\
MO &6.75 &29.00 &9.00 &5.00 &67.44 &20.93 &11.63 &0.16 \\
MS &6.42 &17.00 &4.00 &6.00 &62.96 &14.81 &22.22 &0.24 \\
MT &1.02 &18.00 &15.00 &17.00 &36.00 &30.00 &34.00 &0.02 \\
NC &13.30 &57.00 &39.00 &13.00 &52.29 &35.78 &11.93 &0.12 \\
ND &9.29 &30.00 &13.00 &7.00 &60.00 &26.00 &14.00 &0.19 \\
NE &6.85 &18.00 &3.00 &6.00 &66.67 &11.11 &22.22 &0.25 \\
NH &4.89 &27.00 &18.00 &10.00 &49.09 &32.73 &18.18 &0.09 \\
NJ &12.93 &33.00 &10.00 &12.00 &60.00 &18.18 &21.82 &0.24 \\
NM &12.50 &46.00 &22.00 &21.00 &51.69 &24.72 &23.60 &0.14 \\
NV &21.51 &60.00 &29.00 &29.00 &50.85 &24.58 &24.58 &0.18 \\
NY &12.22 &29.00 &14.00 &13.00 &51.79 &25.00 &23.21 &0.22 \\
OH &7.05 &23.00 &13.00 &16.00 &44.23 &25.00 &30.77 &0.14 \\
OK &22.32 &64.00 &28.00 &19.00 &57.66 &25.23 &17.12 &0.20 \\
OR &7.12 &21.00 &10.00 &10.00 &51.22 &24.39 &24.39 &0.17 \\
PA &11.42 &40.00 &17.00 &24.00 &49.38 &20.99 &29.63 &0.14 \\
RI &11.02 &26.00 &7.00 &6.00 &66.67 &17.95 &15.38 &0.28 \\
SC &3.17 &15.00 &10.00 &9.00 &44.12 &29.41 &26.47 &0.09 \\
SD &4.21 &17.00 &7.00 &2.00 &65.38 &26.92 &7.69 &0.16 \\
TN &11.11 &37.00 &19.00 &12.00 &54.41 &27.94 &17.65 &0.16 \\
TX &34.05 &66.00 &11.00 &16.00 &70.97 &11.83 &17.20 &0.37 \\
UT &6.64 &19.00 &8.00 &3.00 &63.33 &26.67 &10.00 &0.22 \\
VA &8.95 &36.00 &20.00 &12.00 &52.94 &29.41 &17.65 &0.13 \\
VT &5.01 &20.00 &7.00 &7.00 &58.82 &20.59 &20.59 &0.15 \\
WA &15.93 &43.00 &18.00 &12.00 &58.90 &24.66 &16.44 &0.22 \\
WI &1.60 &13.00 &16.00 &4.00 &39.39 &48.48 &12.12 &0.05 \\
WV &12.78 &39.00 &17.00 &18.00 &52.70 &22.97 &24.32 &0.17 \\
WY &14.89 &41.00 &14.00 &12.00 &61.19 &20.90 &17.91 &0.22 \\
\bottomrule
\end{tabular}
\end{table*}

\begin{table*}\centering
\caption{LLM Question + Answer Sentiment Analysis}\label{tab:llm_sentiment_analysis_qa}
\scriptsize
\begin{tabular}{lrrrrrrrrr}\toprule
STATE &Total Compound &Positive &Negative &Neutral &Positive \% &Negative \% &Neutral \% &Average Sentiment Score \\\midrule
AK &0.53 &22.00 &25.00 &8.00 &40.00 &45.45 &14.54 &0.01 \\
AL &11.14 &27.00 &7.00 &8.00 &64.29 &16.67 &19.05 &0.27 \\
AR &1.71 &10.00 &7.00 &5.00 &45.45 &31.82 &22.73 &0.08 \\
AZ &13.47 &26.00 &5.00 &4.00 &74.29 &14.29 &11.43 &0.39 \\
CA &26.86 &55.00 &12.00 &13.00 &68.75 &15.00 &16.25 &0.34 \\
CO &14.06 &30.00 &7.00 &8.00 &66.67 &15.56 &17.78 &0.31 \\
CT &6.14 &28.00 &18.00 &13.00 &47.46 &30.51 &22.03 &0.10 \\
DE &31.01 &61.00 &20.00 &23.00 &58.65 &19.23 &22.11 &0.30 \\
FL &41.48 &90.00 &26.00 &16.00 &68.18 &19.70 &12.12 &0.31 \\
GA &-0.68 &31.00 &31.00 &17.00 &39.24 &39.24 &21.52 &-0.01 \\
HI &13.36 &37.00 &15.00 &23.00 &49.33 &20.00 &30.67 &0.18 \\
IA &5.68 &29.00 &18.00 &11.00 &50.00 &31.03 &18.97 &0.10 \\
ID &2.51 &22.00 &14.00 &16.00 &42.31 &26.92 &30.77 &0.05 \\
IL &25.28 &64.00 &11.00 &15.00 &71.11 &12.22 &16.67 &0.28 \\
IN &12.17 &42.00 &17.00 &21.00 &52.50 &21.25 &26.25 &0.15 \\
KS &8.68 &27.00 &9.00 &9.00 &60.00 &20.00 &20.00 &0.19 \\
KY &5.24 &26.00 &16.00 &15.00 &45.61 &28.07 &26.32 &0.09 \\
LA &2.74 &48.00 &49.00 &17.00 &42.10 &42.98 &14.91 &0.02 \\
MA &15.63 &46.00 &19.00 &25.00 &51.11 &21.11 &27.78 &0.17 \\
MD &12.07 &25.00 &10.00 &10.00 &55.56 &22.22 &22.22 &0.27 \\
ME &4.45 &28.00 &20.00 &8.00 &50.00 &35.71 &14.29 &0.08 \\
MI &18.94 &90.00 &46.00 &21.00 &57.33 &29.30 &13.38 &0.12 \\
MN &3.59 &34.00 &29.00 &16.00 &43.04 &36.71 &20.25 &0.05 \\
MO &5.56 &25.00 &13.00 &5.00 &58.14 &30.23 &11.63 &0.13 \\
MS &6.21 &17.00 &6.00 &4.00 &62.96 &22.22 &14.81 &0.23 \\
MT &-1.76 &17.00 &18.00 &15.00 &34.00 &36.00 &30.00 &-0.03 \\
NC &10.81 &57.00 &42.00 &10.00 &52.29 &38.53 &9.17 &0.10 \\
ND &8.56 &30.00 &13.00 &7.00 &60.00 &26.00 &14.00 &0.17 \\
NE &7.20 &19.00 &3.00 &5.00 &70.37 &11.11 &18.52 &0.27 \\
NH &5.62 &29.00 &18.00 &8.00 &52.73 &32.73 &14.54 &0.10 \\
NJ &11.15 &35.00 &12.00 &8.00 &63.64 &21.82 &14.54 &0.20 \\
NM &11.74 &49.00 &24.00 &16.00 &55.06 &26.97 &17.98 &0.13 \\
NV &19.94 &61.00 &31.00 &26.00 &51.70 &26.27 &22.03 &0.17 \\
NY &13.70 &31.00 &12.00 &13.00 &55.36 &21.43 &23.21 &0.25 \\
OH &8.06 &26.00 &14.00 &12.00 &50.00 &26.92 &23.08 &0.16 \\
OK &21.86 &66.00 &30.00 &15.00 &59.46 &27.03 &13.51 &0.20 \\
OR &6.91 &21.00 &11.00 &9.00 &51.22 &26.83 &21.95 &0.17 \\
PA &7.60 &41.00 &25.00 &15.00 &50.62 &30.86 &18.52 &0.09 \\
RI &10.43 &25.00 &8.00 &6.00 &64.10 &20.51 &15.38 &0.27 \\
SC &3.65 &16.00 &10.00 &8.00 &47.06 &29.41 &23.53 &0.11 \\
SD &3.76 &17.00 &7.00 &2.00 &65.39 &26.92 &7.69 &0.15 \\
TN &11.26 &39.00 &21.00 &8.00 &57.35 &30.88 &11.77 &0.17 \\
TX &38.36 &68.00 &11.00 &14.00 &73.12 &11.83 &15.05 &0.41 \\
UT &6.82 &19.00 &8.00 &3.00 &63.33 &26.67 &10.00 &0.23 \\
VA &5.54 &35.00 &23.00 &10.00 &51.47 &33.82 &14.71 &0.08 \\
VT &5.71 &20.00 &7.00 &7.00 &58.82 &20.59 &20.59 &0.17 \\
WA &16.36 &43.00 &18.00 &12.00 &58.90 &24.66 &16.44 &0.22 \\
WI &-0.25 &14.00 &15.00 &4.00 &42.42 &45.45 &12.12 &-0.01 \\
WV &12.58 &39.00 &21.00 &14.00 &52.70 &28.38 &18.92 &0.17 \\
WY &13.94 &40.00 &16.00 &11.00 &59.70 &23.88 &16.42 &0.21 \\
\bottomrule
\end{tabular}
\end{table*}

\end{document}